\documentclass{aa}
\usepackage[varg]{txfonts}
\usepackage{psfig,graphicx,natbib, gensymb}
\usepackage{color}

\begin{document}

\title{Exploring the stellar surface phenomena of WASP-52 and HAT-P-30 with ESPRESSO\thanks{Based on observations made at ESO's VLT (ESO Paranal Observatory, Chile) under ESO programmes 0102.C-0493 and 0102.D-0789.}}
\titlerunning{Stellar surface phenomena of WASP-52 and HAT-P-30}
\authorrunning{Cegla et al.}

\author{H.~M. Cegla\inst{1,2,3}\fnmsep\thanks{UKRI Future Leaders Fellow}
	\and N. Roguet-Kern\inst{1}
	\and M. Lendl\inst{1}
	\and B. Akinsanmi\inst{1}
	\and J. McCormac\inst{2,3}
	\and M. Oshagh\inst{4,5}
	\and P.~J. Wheatley\inst{2,3}
	\and G. Chen\inst{6}
	\and R. Allart\inst{7}
	\and A. Mortier\inst{8}
	\and V. Bourrier\inst{1}
	\and N. Buchschacher\inst{1}
	\and C. Lovis\inst{1}
	\and D. Sosnowska\inst{1}
	\and S. Sulis\inst{9}
	\and O. Turner\inst{1}
	\and N. Casasayas-Barris\inst{4,5}
	\and E.~Palle\inst{4,5}
	\and F. Yan\inst{10}
	\and M.~R. Burleigh\inst{11}
	\and S.~L. Casewell\inst{11}
	\and M.~R. Goad\inst{11}
	\and F. Hawthorn\inst{2,3}
	\and A. Wyttenbach\inst{1}
} 

\offprints{H. M. Cegla, \email{h.cegla@warwick.ac.uk}}

\institute{Observatoire Astronomique de l'Universit\'e de Gen\`eve, Chemin Pegasi 51b, CH-1290 Versoix, Switzerland
\and
Physics Department, University of Warwick, Coventry CV4 7AL, United Kingdom
  \and Centre for Exoplanets and Habitability, University of Warwick, Coventry CV4 7AL, UK
  \and Instituto de Astrof\'isica de Canarias (IAC), 38205 La Laguna, Tenerife, Spain
  \and Deptartamento de Astrof\'isica, Universidad de La Laguna (ULL), 38206 La Laguna, Tenerife, Spain
  \and CAS Key Laboratory of Planetary Sciences, Purple Mountain Observatory, Chinese Academy of Sciences, Nanjing 210023, China
  \and Department of Physics, and Institute for Research on Exoplanets, Universit\'e de Montr\'eal, Montr\'eal, H3T 1J4, Canada
  \and School of Physics \& Astronomy, University of Birmingham, Edgbaston, Birmingham, B15 2TT, UK
  \and Aix Marseille Univ, CNRS, CNES, LAM, 38 rue Frédéric Joliot-Curie, 13388 Marseille, France
  \and Department of Astronomy, University of Science and Technology of China, Hefei 230026, China
  \and Department of Physics and Astronomy, University of Leicester, University Road, Leicester, LE1 7RH, UK
}

\date{Received X X XXXX / Accepted X X XXXX}
\abstract{We analyse spectroscopic and photometric transits of the hot Jupiters WASP-52~b and HAT-P30~b obtained with ESPRESSO, Eulercam and NGTS for both targets, and additional TESS data for HAT-P-30. Our goal is to update the system parameters and refine our knowledge of the host star surfaces. For WASP-52, the companion planet has occulted starspots in the past, and as such our aim was to use the reloaded Rossiter–McLaughlin technique to directly probe its starspot properties. Unfortunately, we find no evidence for starspot occultations in the datasets herein. Additionally, we searched for stellar surface differential rotation (DR) and any centre-to-limb variation (CLV) due to convection, but return a null detection of both. This is unsurprising for WASP-52, given its relatively cool temperature, high magnetic activity (which leads to lower CLV), and projected obliquity near 0$\degree$ (meaning the transit chord is less likely to cross several stellar latitudes). For HAT-P-30, this result was more surprising given its hotter effective temperature, lower magnetic field, and high projected obliquity (near 70$\degree$). To explore the reasons behind the null DR and CLV detection for HAT-P-30, we simulated a variety of scenarios. We find that either the CLV present on HAT-P-30 is below the solar level or the presence of DR prevents a CLV detection given the precision of the data herein. A careful treatment of both DR and CLV is required, especially for systems with high impact factors, due to potential degeneracies between the two. Future observations and/or a sophisticated treatment of the red noise present in the data (likely due to granulation) is required to refine the DR and CLV for these particular systems; such observations would also present another opportunity to try to examine starspots on WASP-52.    
}


\maketitle

\section{Introduction}
\label{sec:intro}
Transiting exoplanets have the power to open a new and unique observational window on stellar astrophysics, which in turn can feed back into exoplanet confirmation and characterisation techniques. The surfaces of Sun-like stars have numerous inhomogeneities (e.g., but not limited to, magnetoconvection, pressure-mode oscillations, stellar differential rotation, spots, faculae/plage) that contaminate planetary measurements. For example, stellar surface phenomena alter the observed stellar absorption line profiles and introduce asymmetries that may be mistaken for wholesale Doppler shifts that can mask or mimic the Doppler reflex motion induced by a planetary companion \citep{saar97} -- even solar-level convection/granulation and p-mode oscillations alone are large enough to completely disguise the Doppler wobble of an Earth twin \citep[][and references therein]{cegla19a}. Starspots, faculae/plage, and limb-dependent convective variations have all been shown to contaminate planet atmosphere interpretations if treated incorrectly, even introducing false-positive signatures \citep[e.g.][and references therein]{mccullough14,oshagh14,bruno20, casasayas-barris20, chen20}. Ignoring these effects can also bias our measurements of star-planet system geometries \citep[][]{czesla15,cegla16a, oshagh16, oshagh18}, as well as our stellar rotation interpretations \citep[e.g.][]{triaud09, triaud15,cegla16b}. Moreover, as stellar differential rotation is believed to play a key role in the dynamo process that generates magnetic fields, systematically neglecting it in exoplanet observations may also bias our understanding of magnetic activity as a whole. The ability to spatially resolve our Sun provides a wealth of information; however, transiting planets allow the unique opportunity to spatially resolve and explore the physics at play in distant stars across a variety of spectral types.

One such technique that harnesses this power of transiting exoplanets is the reloaded Rossiter–McLaughlin (RRM). In its first application, using data from HARPS (High Accuracy Radial velocity Planet Searcher), \cite{cegla16b} were able to put constraints on the stellar surface differential rotation of HD~189733, self-consistently derive the system's 3D obliquity, and place limits on the centre-to-limb variation in the net convective velocities. However, HD~189733 is a slowly rotating, magnetically active K dwarf, with an aligned planetary companion; these combined aspects mean that while this system's brightness lends itself to high precision radial velocities (RVs), the overall amplitude of the differential rotation and convective centre-to-limb variations is low and therefore difficult to measure \citep{Roguet-Kern22}. Moreover, while this host star is magnetically active, the planet does not transit these regions and therefore cannot be used to directly probe them. Similar constraints were true for other RRM analyses \citep{bourrier17, bourrier18, bourrier22}; however, we note that their use of the RRM did allow them to make significant advances in the star-planet obliquity measurements, including the first such measurements for an M dwarf. In a recent study by \cite{doyle22}, the authors were able to put stronger constraints on the convective centre-to-limb variations of WASP-166 and even pull out a tentative detection using the RRM technique. The convective centre-to-limb variations of specific stellar lines were also retrieved using transiting exoplanets as probes of the stellar surfaces of HD~209458 and HD~189733 \citep{dravins17, dravins18}; for a more in-depth discussion on future observational prospects, we advise the reader to see \cite{dravins21}. Similarly, for a more in-depth discussion on stellar surface differential rotation across spectral types, using different techniques, we advise the reader to see \cite{balona16} and references therein.

Nonetheless, we can advance the stellar characterisation capability via the RRM if we optimise the instrument and/or telescope choice and the target selection. Higher precision spectrographs and larger telescope diameters naturally improve the RV precision of individual measurements, while higher resolution spectrographs allow us to further resolve any potential centre-to-limb variations that arise from changes in the stellar line profile shapes. On the other hand, stars with a higher effective temperature, lower magnetic activity, and/or a later stage of evolution can have a higher amplitude centre-to-limb variation due to convection. Similarly, stars with equivalent differential rotation, but a higher overall stellar rotation have a larger amplitude effect that is easier to measure. There are also some star-planet geometries that are more favourable than others; for example, geometries where the planet crosses more stellar latitudes may make it easier to detect the surface differential rotation \citep{serrano20,Roguet-Kern22}. Additionally, if the planet occults an active region, then we can isolate the starlight from these areas and probe the impact of the magnetic field directly. 

It is for the reasons above that we analyse transit observations of WASP-52~b and HAT-P-30~b\footnote{also known as WASP-51~b}, taken with the ESPRESSO spectrograph on the Very Large Telescope (VLT) as part of programmes 0102.D-0789 (PI: H.~M. Cegla) and 0102.C-0493 (PI: G. Chen). WASP-52 is an interesting system as its hot Jupiter is known to occult regions of increased magnetic activity on its relatively bright host, including evidence for both spots and faculae/plage \citep{kirk16,mancini17, bruno18, ozturk19}. If we can isolate the local cross-correlation function (CCF) from such an active region, with the reloaded RM, then we can compare its shape and net convective blueshift with regions from the quiescent photosphere, as well as predictions from simulations. Herein, we combine the spectroscopic transits with both simultaneous and long-term photometric monitoring of WASP-52 to identify potential active regions. Previous RM observations of WASP-52 also indicated that this system might have a small misalignment (24$^{+17}_{-9}{}\degree$; \citealt{hebrard12}), which would make detecting stellar differential rotation slightly easier; however, the ESPRESSO observations clearly show this system has a projected alignment near zero \citep[][and this work herein]{chen20}, in line with the photometric analysis of \cite{mancini17}. Nonetheless, the increased resolution, RV precision and collecting power of ESPRESSO at the VLT means this target still warrants investigation in regards to the detection of stellar surface differential rotation, as well as centre-to-limb variations in the net convective blueshift. On the other hand, HAT-P-30 was identified as an ideal candidate for probing stellar surface differential rotation as the previously determined projected obliquity of its hot Jupiter companion is relatively large, $73.5 \pm 9.0\degree$ \citep{johnson11}, meaning there is a high likelihood that the planet will occult multiple latitudes. Additionally, HAT-P-30's high effective temperature ($\sim$6300~K) and evidence for low magnetic activity (based on photometric monitoring and its $\log R'_{HK}$) means the centre-to-limb variation in its net convective blueshift should be higher and more easily detectable. 

Both WASP-52 and HAT-P-30 rotate relatively slowly, $\sim$3~km~s$^{-1}$ and $\sim$2~km~s$^{-1}$, respectively; this means the impact of stellar surface differential rotation (DR) will be subtle and similar in scale to the expected centre-to-limb variations (CLV) in the net convective blueshift. As a result, we also perform injection/recovery tests for these signatures using model star observations, combined with local RV uncertainties based on (and extrapolated from) the empirical data. 

In Section~\ref{sec:obs}, we detail the spectroscopic and photometric observations taken as a part of this campaign. The transit light curves and analysis of the long-term photometry is described in Section~\ref{sec:LC}, while the reloaded RM analysis is performed in Section~\ref{sec:RM}. The simulated DR and CLV injection/recovery tests can be found in Section~\ref{sec:sim}. We summarise and conclude in Section~\ref{sec:conc}.  

\begin{center}
\begin{table*}
 \caption{Summary of the ESPRESSO observations, including the number of observations per night (with the in- and out-of-transit breakdown in brackets alongside those in-transit that passed the quality threshold) and exposure time. S/N is the average from order 103, i.e. near $\sim$550~nm.}
 \centering
 \begin{tabular}{ccccc}
    \hline
     \hline
   Target& Night & N$_{\rm{obs}}$ (in [pass] /out) & t$_{\rm{exp}}$ & S/N  \\
   \hline
   WASP-52$^{\ast}$& 31-10-2018 & 24 (11 [11] / 13) & 500~s & 27 \\  
   WASP-52$^{\dagger}$& 07-11-2018 & 18 (8 [7] / 10) & 800~s & 31 \\ 
   WASP-52$^{\dagger}$& 14-11-2018 & 15 (8 [7] /7) & 800~s & 35 \\ 
   HAT-P-30$^{\ast}$& 10-01-2019 & 53 (31 [25] / 22) & 200~s & 38 \\
   HAT-P-30$^{\ast}$& 27-03-2019 & 52 (30 [23] /22) & 200~s & 35 \\
   \hline
   \footnotesize{$^{\ast}$Run ID: 0102.D-0789} \\
   \footnotesize{$^{\dagger}$Run ID: 0102.C-0493} \\
 \end{tabular}
  \label{tab:espresso} 
 \end{table*} 
 \end{center}
 \vspace{-20pt}

\section{Observations}
\label{sec:obs}
We observed simultaneous spectroscopic and photometric transits of two hot Jupiters: WASP-52~b and HAT-P-30~b (also known as WASP-51~b; to avoid confusion, we use the HAT nomenclature hereafter). 

\subsection{Spectroscopic observations}
Spectroscopic transits were observed with the ESPRESSO (Echelle SPectrograph for Rocky Exoplanets and Stable Spectroscopic Observations) spectrograph in single UT mode on one of the 8~m telescopes of the VLT (Very Large Telescope) at the ESO Paranal as part of programmes 0102.D-0789 (PI: H.~M. Cegla) and 0102.C-0493 (PI: G. Chen). Successful transit observations of WASP-52~b, with ESPRESSO, were taken on October 31st, November 7th, and November 14th 2018 (an attempt to observe a transit on October 3rd was unsuccessful due to a failure in the telescope control system). Transits of HAT-P-30~b were observed with ESPRESSO on January 10th and March 27th 2019. All ESPRESSO observations employed the 2x1 binning in slow readout, high resolution mode (R$\sim$140 000); further details can be found in the summary in Table~\ref{tab:espresso}. Each set of observations has approximately a 1~hr baseline before and after transit; all observations were reduced with version 2 of the ESPRESSO Data Reduction Software \footnote{ftp://ftp.eso.org/pub/dfs/pipelines/espresso/espdr-reflex-tutorial-2.2.1.pdf}. 

\begin{table*}
\caption[]{\label{tab:phot}Summary of photometric observations included in this work. The nomenclature for the baseline models, $p^i(\mathrm{par_1})$, refers to an \emph{i-th} order polynomial in parameter $\mathrm{par_1}$, with $t$, $\mathit{fwhm}$ and $\mathit{xy}$ referring to time, full-width at half maximum and coordinate shifts, respectively and $\mathit{off}$ denoting a constant offset. The $\sigma_w$ values refer to supplementary white noise resulting from the CONAN analysis of each light curve, added to the uncertainties in quadrature; see Section~{\ref{sec:LC}}.}
\begin{center}
\begin{tabular}{cccccc}

    \hline
    \hline
      Date & Instrument & Bandpass & Baseline model & White noise $\sigma_w$ [ppm] & RMS (2 min) [ppm] \\
    \hline
    \multicolumn{4}{l}{WASP-52} \\
      31-10-2018 & EulerCam & Geneva V & $p^3(t)+p^1(\mathit{fwhm})$ & 883 & 1298 \\
      31-10-2018 & NGTS & NGTS &  $p^1(t) $ &  7588 & 3483 \\
      01-11-2018 & NGTS & NGTS & $p^1(t) $ & 8729 & 4158 \\
      07-11-2018 & EulerCam & Geneva V & $p^1(t)$ & 736 & 1248\\
      07-11-2018 & NGTS & NGTS & $p^1(t) $ & 8865 & 3893 \\
      14-11-2018 & NGTS & NGTS & $p^1(t) $ & 7080 & 3512 \\
      
    \hline
    \multicolumn{4}{l}{HAT-P-30} \\
      10-01-2019 & EulerCam & Geneva V & $p^2(t)$ & 528 & 892  \\
      10-01-2019 & NGTS & NGTS & $p^1(t)$ & 867 & 2241 \\
      27-03-2019 & EulerCam & Geneva V & $p^1(t)+p^2(\mathit{xy})$ & 870  & 1173 \\
      27-03-2019 & NGTS & NGTS & $p^1(t)$ & 526 & 2385 \\
      January 2019 & TESS & TESS & $\mathit{off}$ & 0 & 977 \\
    \hline  
    \hline
  \end{tabular}
\end{center}
\label{tab:lc_summary}
\end{table*}

\subsection{Photometric observations}
We obtained photometric transit observations of both WASP-52 and HAT-P-30 with the goal of revising planetary parameters, measuring a precise ephemeris of the targets, and searching for the signatures of spot-crossing events in simultaneous observations. To do so, we used both EulerCam at the 1.2m Euler telescope \citep{Lendl12} and the Next Generation Transit Survey (NGTS) \citep{Wheatley18}; the latter consists of twelve 20~cm telescopes located at Paranal observatory. All ground-based photometry was obtained using relative aperture photometry with iterative selection of stable reference stars and aperture sizes. For HAT-P-30, we also make use of data provided by the photometric TESS satellite \citep{Ricker15}. 

\emph{WASP-52:} We obtained photometric transit observations in parallel with the first two ESPRESSO spectroscopic observations with EulerCam using a Geneva V band filter \citep{Rufener88}. The choice of a comparably blue band was motivated by the aim of increasing sensitivity to spot-crossing events manifesting themselves as bumps in the transit light curve. With NGTS (using a single 20~cm telescope), we obtained parallel observations to all three ESPRESSO observations, plus two additional transit observations on 2018-09-12 and 2018-10-31. NGTS is equipped with a single broad optical filter with high transmission between 500 and 900~nm \citep{Wheatley18}.

In the interest of refining/updating the stellar rotation period for WASP-52, we also performed long-term photometric monitoring for this target. The discovery paper reported a stellar rotation period near $\sim$12~d \citep{hebrard12} based on their $v \sin i_{\star}$ and stellar radius measurements, assuming the rotation axis of the star was aligned with the orbital plane of the planet; however, they also reported a $\sim$16~d periodicity observed in two seasons of WASP data. Later reports from \cite{louden17, bruno20, rosich20} favour a stellar rotation period closer to $\sim$16-18~d. To fully sample these periodicities and explore the stellar variability present near our transit observations, we obtained 79 nights of photometric monitoring with NGTS between 30 August 2018 and 1 December 2018. An average of 50 10 second images were obtained each night and combined into a single, high precision photometric measurement.

\emph{HAT-P-30:} We obtained photometric transit observations with both, EulerCam and NGTS in parallel to both ESPRESSO observations. As for WASP-52, a Geneva V band filter \citep{Rufener88} was used for EulerCam, while NGTS observations were made using a single 20~cm telescope with a broad optical bandpass \citep{Wheatley18}. We also include TESS observations of HAT-P-30 obtained during its Sector 7. We made use of the TESS Science Processing Operations Center \citep{Jenkins16} 2-min cadence PDC light curve \citep{Smith12,Stumpe14}, which we additionally filtered for low-frequency trends using a 6-hour boxcar while masking the transits of planet b. All light curves are summarised in Table \ref{tab:phot}.

\begin{center}
\begin{figure}[t!]
\centering
\includegraphics[width=\linewidth]{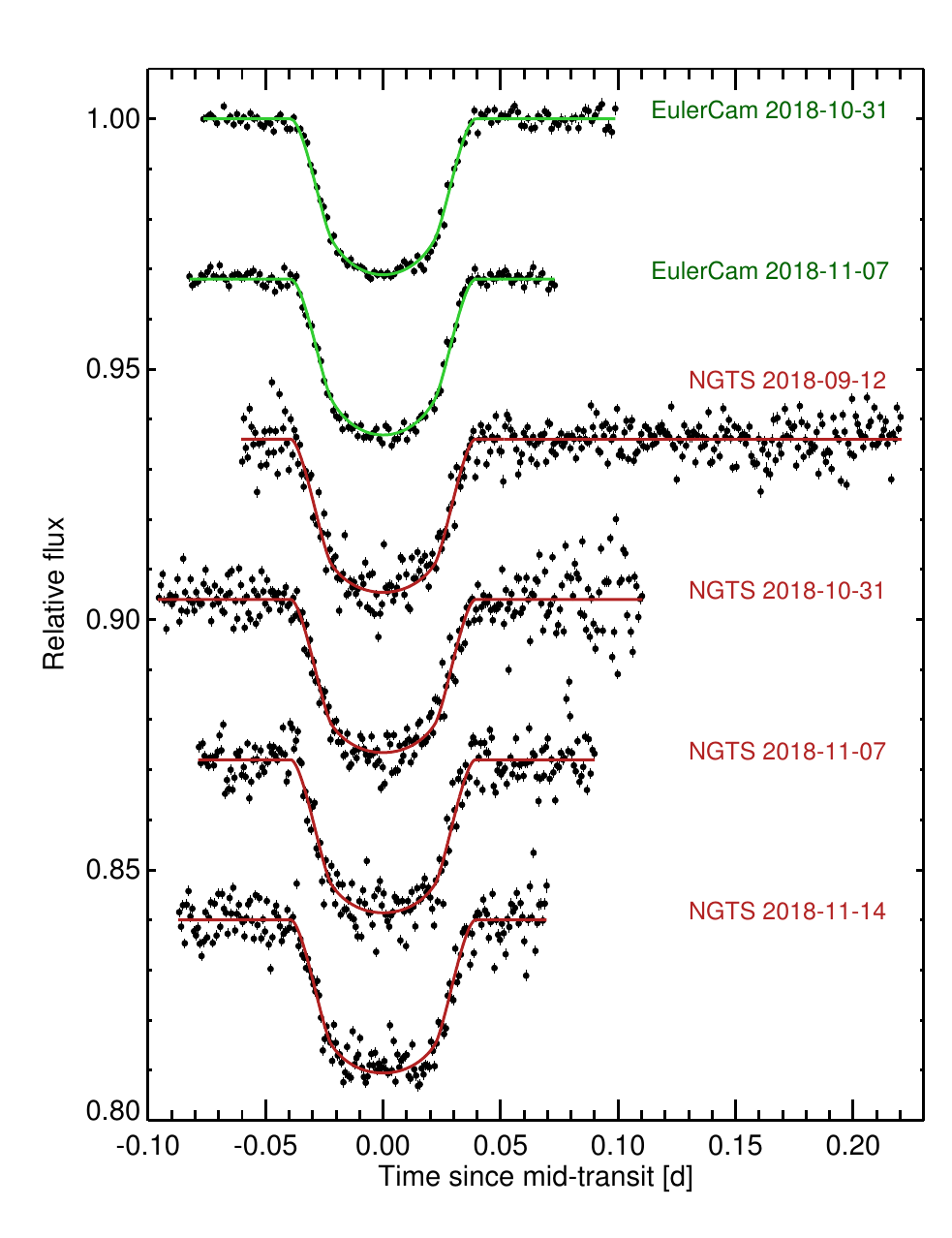}
\caption[]{\label{fig:photW52} Ground-based transit photometry of WASP-52 together with the best-fit models (solid curves). The data have been baseline corrected and binned into 2-minute intervals.
} 
\label{fig:}
\end{figure}
\end{center}

\begin{center}
\begin{figure}[t!]
\centering
\includegraphics[width=\linewidth]{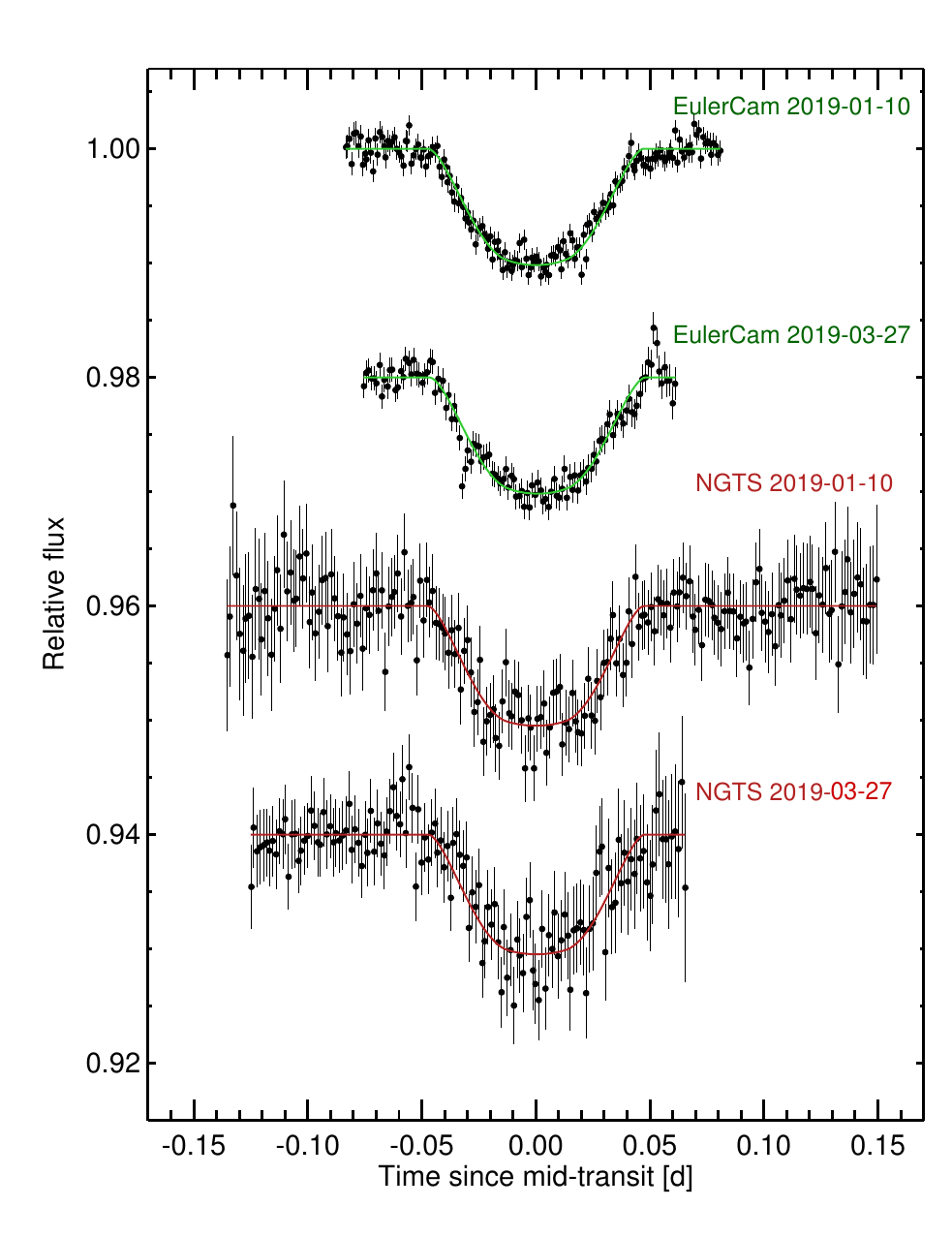}
\caption[]{\label{fig:photHAT30ground} Ground-based transit photometry of HAT-P-30 together with the best-fit models (solid curves). The data have been baseline corrected and binned into 2-minute intervals.
} 
\label{fig:}
\end{figure}
\end{center}

\begin{center}
\begin{figure}[t!]
\centering
\includegraphics[width=\linewidth]{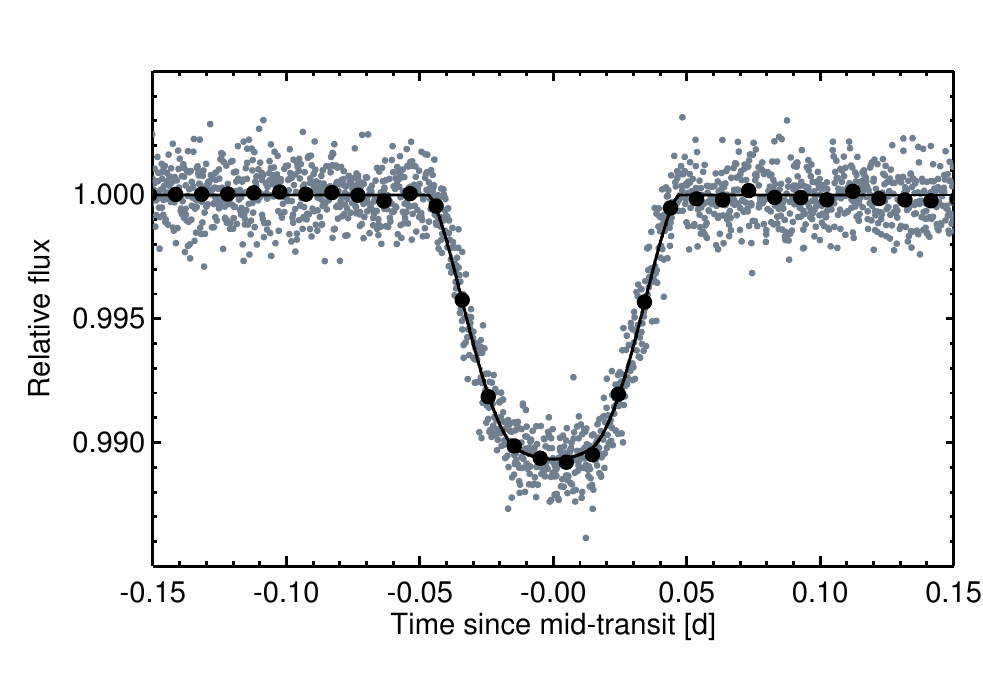}
\caption[]{\label{fig:photHAT30tess} TESS transit photometry of HAT-P-30 together with the best-fit models. The data have been phase-folded, with data binned into 14-minute intervals shown in black.
} 
\label{fig:}
\end{figure}
\end{center}

\begin{center}
\begin{figure*}[t!]
\centering
\includegraphics[scale=0.25]{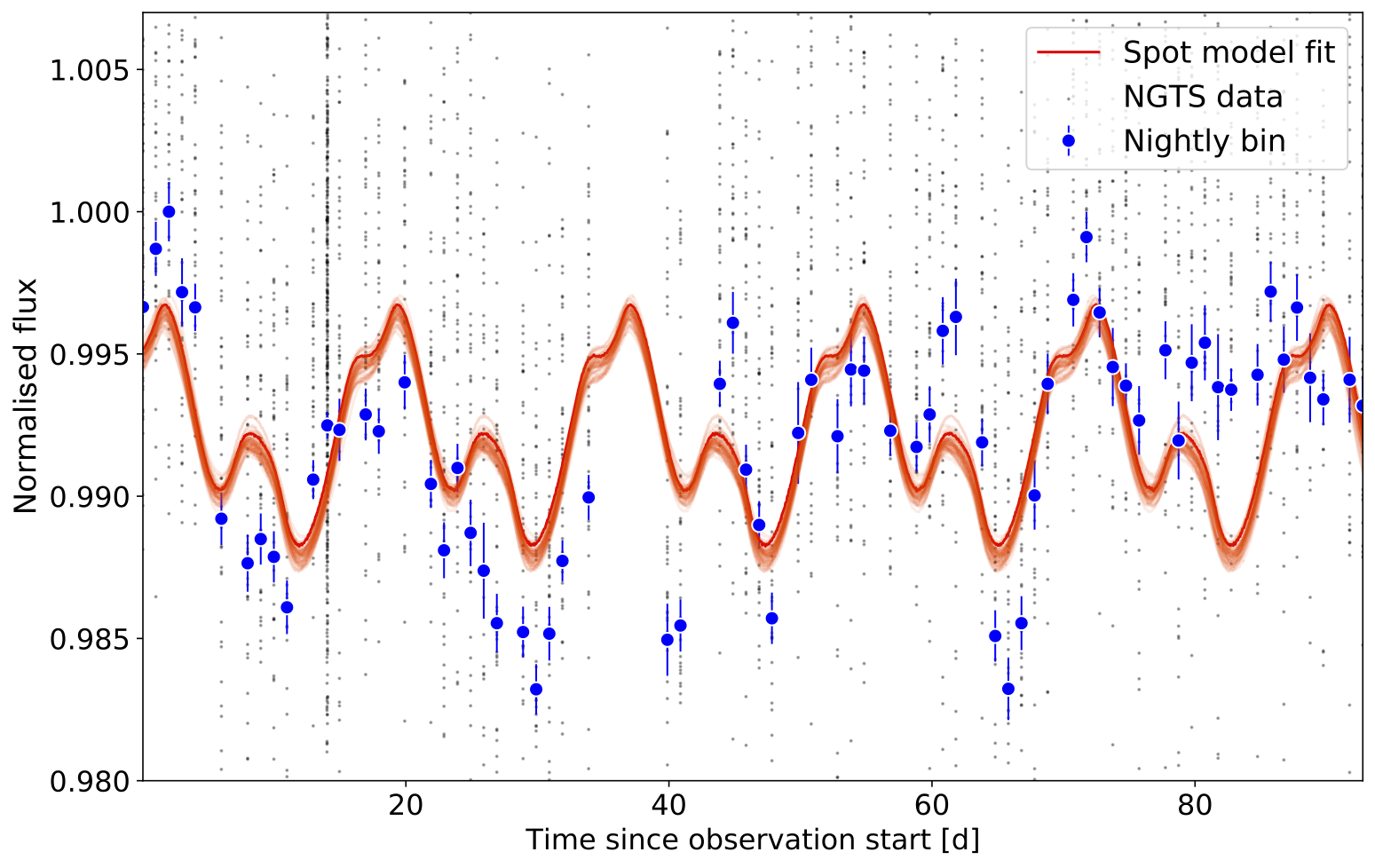}
\caption[]{Three-spot model fit to the NGTS photometry of WASP-52. The data is binned per night and the red curves show random samples from the posterior distribution; spot evolution, which is unaccounted for in \emph{SOAP}, is the most likely reason for the divergence in fit.} 
\label{fig:wasp-52_soap_spot_fit}
\end{figure*}
\end{center}

\section{Light curve modelling}
\label{sec:LC}

We used the transit light curves described above to derive updated  parameters and ephemerides for the WASP-52 and HAT-P-30 systems. To do so we used a Markov chain Monte Carlo (MCMC) approach implemented in the \emph{CONAN} package \citep{Lendl20a}. We assumed stellar parameters from \citet{johnson11} and \citet{mancini17} and radial-velocity amplitudes from \citet{johnson11} and \citet{hebrard12} for HAT-P-30 and WASP-52, respectively, and assumed circular orbits following the original RV analysis (which is sensible given their short periods). We further assumed quadratic limb-darkening parameters, fixed to values derived with the \emph{limbdark} tool \citep{Espinoza15} for each passband. A wide range of parametric baseline models were tested to fit systematic trends of each light curve, however low-order polynomials were found to be adequate via comparison of the Bayes factor \citep[e.g.][]{Carlin09}. The baseline models used are listed in Table \ref{tab:phot}, and the light curves, together with their best-fit models are shown in Figures \ref{fig:photW52}, \ref{fig:photHAT30ground}, and \ref{fig:photHAT30tess}. We allowed for additional white noise in each light curve by independently requiring each data set to produce a reduced $\chi^2$ of unity and increasing the uncertainties to reach this condition. A Gelman \& Rubin \citep{Gelman92} test confirmed good convergence of the MCMC chains. The updated parameters are given in Tables \ref{tab:wasp52_param} and \ref{tab:hatp30_param} for WASP-52 and HAT-P-30, respectively.

From \cite{mancini17}, we expect anomalies from spot occultations for WASP-52~b to be of the order of $\sim$0.005-0.01 in normalised flux (e.g., see their Figures 1-3) and to occur over a $\sim$40 minute duration. We note that the 2-minute WASP-52 data from EulerCam has an RMS around 0.001 in normalised flux (see Table~\ref{tab:lc_summary}); as such, we should be able to detect any potential spot crossings with high significance. Unfortunately, as shown in Figure~\ref{fig:photW52}, we do not find any significant evidence for spot occultations during the WASP-52~b transits herein. 

\begin{table}[t!]
\caption[]{Star-planet parameters for the WASP-52 system}
\begin{center}
\begin{tabular}{ccc}

    \hline
    \hline
      Par. & Value & Reference \\
    \hline  
	$T_{0}$ & 2458374.60896530$^{+0.00037832}_{-0.00028912}$ d & This work \\
	$P$ & 1.74977083$^{+0.00001028}_{-0.00001200}$ d & This work\\
	$i_p$ & 1.48573310$^{+0.00377061}_{-0.00326637}$ & This work \\
	$R_p$ & 0.16739970$^{+0.00109760}_{-0.00113988}$ R$_{\star}$ & This work\\
	$a/R_{\star}$ & 7.16832287$^{+0.11414292}_{-0.10113834}$ & This work \\
	$u_1$ & 0.69880 & This work \\
	$u_2$ & 0.08777 & This work \\
	
    $K$ & 0.0843 $\pm$ 0.0030 km s$^{-1}$ & \citealt{hebrard12} \\
	$e$ & 0 & \citealt{hebrard12} \\
	$\omega$ & 90$\degree$ & \citealt{hebrard12} \\
	
	T$_{eff}$ & 5014 $\pm$ 41 K & \citealt{chen20}\\
	log $g$ & 4.48 $\pm$ 0.08 & \citealt{chen20}\\
	V$_{mag}$ & 12.192 $\pm$ 0.069 & \citealt{stassun19}\\
	Fe/H & 0.06 $\pm$ 0.03 & \citealt{chen20}\\
    \hline
  \end{tabular}
\end{center}
\label{tab:wasp52_param}
\end{table}

\begin{table}[t!]
\caption[]{Star-planet parameters for the HAT-P-30 system}
\begin{center}
\begin{tabular}{ccc}

    \hline
    \hline
      Par. & Value & Reference \\
    \hline  
	$T_{0}$ & 2458491.91561223$^{+0.00019437}_{-0.00024875}$ d & This work \\
	$P$ & 2.81057534$^{+0.00003581}_{-0.00003041}$ d & This work\\
	$i_p$ & 1.43941624$^{+0.00257064}_{-0.00231606}$ & This work \\
	$R_p$ & 0.10896401$^{+0.00060343}_{-0.00086416}$ R$_{\star}$ & This work\\
	$a/R_{\star}$ & 6.61749789$^{+0.09170950}_{-0.08749245}$ & This work \\
	$u_1$ & 0.48526013 & This work \\
	$u_2$ & 0.20773733 & This work \\
	
    $K$ & 0.0907 $\pm$ 0.0021 km s$^{-1}$ & \citealt{maciejewski16} \\
	$e$ & 0 &  \citealt{maciejewski16} \\
	$\omega$ & 90$\degree$ &  \citealt{maciejewski16} \\
	
	T$_{eff}$ & 6338 $^{+162}_{-124}$ & \citealt{stassun19}\\
	log $g$ & 4.284$^{+0.028}_{-0.026}$ & \citealt{maciejewski16}\\
	V$_{mag}$ & 10.352 $\pm$ 0.007 & \citealt{stassun19}\\
	Fe/H &  0.130 $\pm$ 0.080 & \citealt{johnson11}\\
    \hline
  \end{tabular}
\end{center}
\label{tab:hatp30_param}
\end{table}

Nonetheless, WASP-52 does show long-term spot modulation, as evidenced in the monitoring from NGTS in Figure~\ref{fig:wasp-52_soap_spot_fit}. We use this long-term photometric monitoring to constrain the potential stellar inclination and rotation rate. We use the\emph{SOAP} code \citep{SOAP_oshagh, SOAP_dumusque, SOAP_akinsanmi18} to fit a spot model to our long-term photometry, shown in Figure~\ref{fig:wasp-52_soap_spot_fit}. We note that the current version of \emph{SOAP} does not consider the evolution of active regions, and as a result the fit is mostly constrained by the first stellar rotation. We assumed different number of spots (2, 3,and 4) and fit for the longitude, latitude and size of each of the spots, alongside the stellar inclination and rotation period. The observations were nightly binned before fitting since spots induce longer-term flux variations. We compared the different models using the Bayes factor and found that the 3-spot model was preferred. From our photometric spot modelling, we find $i_{\star} = 77.1 \pm 15.6{\degree}$ and $P_{rot} = 17.69 \pm 0.12$~d (this $P_{rot}$ is in agreement with the recent analysis by \citealt{rosich20}). 

\section{Reloaded Rossiter-McLaughlin}
\label{sec:RM}
For the reloaded RM method, we create master-out CCFs for each night by co-adding the respective out-of-transit CCFs. Then the master-out and all individual CCFs are continuum normalised and multiplied by a transit light curve (generated with the $batman$ code; \citealt{kreidberg15}) based on the modelling in Section~\ref{sec:LC} to put them on an correct relative scale. We note that we adpot the limb darkening coefficients used in the EulerCam fitting, as its V band filter is a close approximation to the ESPRESSO bandpass; we did test various limb darkening scenarios, with the final results being consistent with 1$\sigma$; however, we caution that for higher precision datasets, the limb darkening choice warrants close consideration as a mismatch can result in systematic biases in the local RVs. In turn, this allows us to directly subtract the individual CCFs from the master-out CCFs, on a night by night basis; for the in-transit observations we isolate the starlight behind the planet, which we refer to as the local CCF, and for the out-of-transit observations the residuals provide information on the precision of the data. We then determine the RV of the local CCF from the centroid of a Gaussian fit. For more details see \cite{cegla16b}.

To model the local RVs, we consider a variety of scenarios for each target: solid body rotation alone (SB), solid body rotation with a centre-to-limb variation in the net convective blueshift (SB+CLV), differential rotation alone (DR), and differential rotation with a CLV (DR+CLV). We follow the coordinate system laid out in \cite{cegla16b}, and similarly assume a solar-like stellar rotation law (but allowing for anti-solar rotation), as well as test both a linear and quadratic limb-dependent polynomial for the CLV. We remind the reader that the disc-integrated net convective blueshift is encoded in the RV of the master-out CCFs, which is removed from the local RVs to account for the systemic velocities and any nightly offsets from atmospheric, instrumental, or slowly evolving stellar variability (relative to the transit timescales, e.g. spots/plage); this subtraction is accounted for in the CLV polynomials within the $c_0$ formulation. 

To perform each model comparison, we utilise the \texttt{emcee} software package \citep{foreman-mackey13} and compute MCMCs with a Gaussian likelihood function. We use 200 walkers, 2000 accepted steps, and following a visual inspection of the chains opt for a burn-in of 500 steps (we also discard a few chains with significantly lower likelihood). We note that for our least constrained cases, we did test using 20000 steps, but found results consistent within 1$\sigma$. We assume uninformative, uniform priors to stay within the physical boundaries and our coordinate system. In particular we constrain the stellar inclination ($i_{\star}$) to 0–180$\degree$, the projected spin-orbit obliquity ($\lambda$) to -180–180$\degree$, and the differential rotation rate ($\alpha$) to -1–1 (note $\alpha$ $<-1$ is physically possible, but we exclude this regime for computational ease and because this parameter space is highly unlikely).  For WASP-52, we use also the long-term photometric monitoring and spot modelling from Section~\ref{sec:LC} to perform additional analyses with Gaussian priors placed on the stellar rotation and inclination. 

Data near the stellar limb have a lower signal-to-noise ratio; this is in part due to increased limb darkening, but largely due to the fact that the planet is only partially on the stellar disc at ingress and egress. To select data of a sufficient quality we set a threshold based off the dispersion of the continuum relative to the contrast (i.e. depth) of the local CCF. The theoretical precision on the local CCF contrast achievable by fitting a Gaussian is $\sigma_{cont} / \sqrt{FWHM}$, where $\sigma_{cont}$ is the dispersion of the local CCF continuum and FWHM is the full-width half maximum expressed in pixel units \citep{allart17}. For our Gaussian fit to perform optimally, the local CCF contrast must be discernible from the surrounding noise. Herein, we exclude data where the local CCF contrast is less than 10 times the precision of the local CCF; this effectively removes data near the stellar limb. Various thresholding factors were tested and each local CCF was examined by eye to confirm the local CCFs were discernible and that the respective Gaussian fits were appropriate.

\begin{figure}[t!]
\includegraphics[trim=0.2cm 10.cm 0.cm 0.cm, clip, scale=0.65]{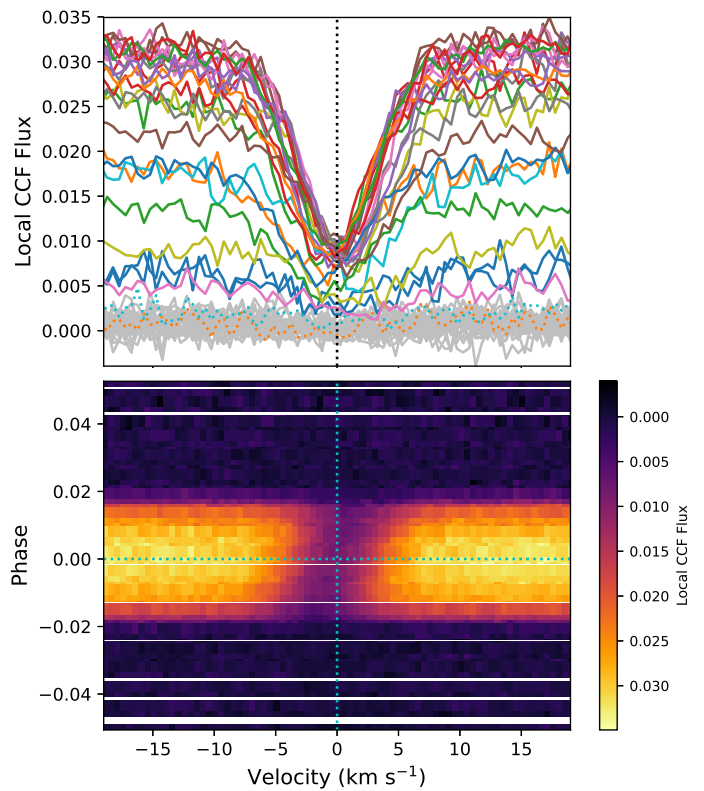}
\includegraphics[trim=0.2cm 0.cm 0.cm 8.7cm, clip, scale=0.65]{WASP-52_loc_cff_phasevelflux}
\caption[]{Local stellar CCFs, behind the transiting planet (i.e. CCF$_{\rm{out}}$ - CCF$_{\rm{in}}$), for WASP-52. Top: plotted in various colours in the stellar rest frame (indicated by a dotted horizontal line); the two in-transit local CCFs that fail the signal-to-noise test are shown as dashed lines. Individual out-of-transit CCFs subtracted from the master CCF$_{\rm{out}}$ are displayed in grey. Bottom: plotted as a function of phase and colour-coded by the flux; a further horizontal dashed line indicates zero phase. } 
\label{fig:wasp52_loc_ccf}
\end{figure}

\subsection{WASP-52}
The local CCFs behind the transiting planet, and along the transit chord, for WASP-52 are shown in Figure~\ref{fig:wasp52_loc_ccf}; the residuals from a subtraction between the master-out and individual out-of-transit CCFs are also shown in grey and provide an indication of the precision. For the three nights of WASP-52 observations, only two local CCFs, very near the stellar limb, fail the aforementioned quality test; they are indicated as dotted lines. Due to the low $v \sin i_{\star}$ of WASP-52 and the noise in the local CCFs, it is hard to discern the stellar rotation by eye, but this becomes easier to see when the local (or residual) CCFs are plotted as a function of phase and colour-coded by flux, as shown in the bottom of Figure~\ref{fig:wasp52_loc_ccf}. 

\begin{center}
\begin{figure}[t!]
\centering
\includegraphics[trim=0.4cm 0.3cm 1.5cm 0.9cm, clip, scale=0.63]{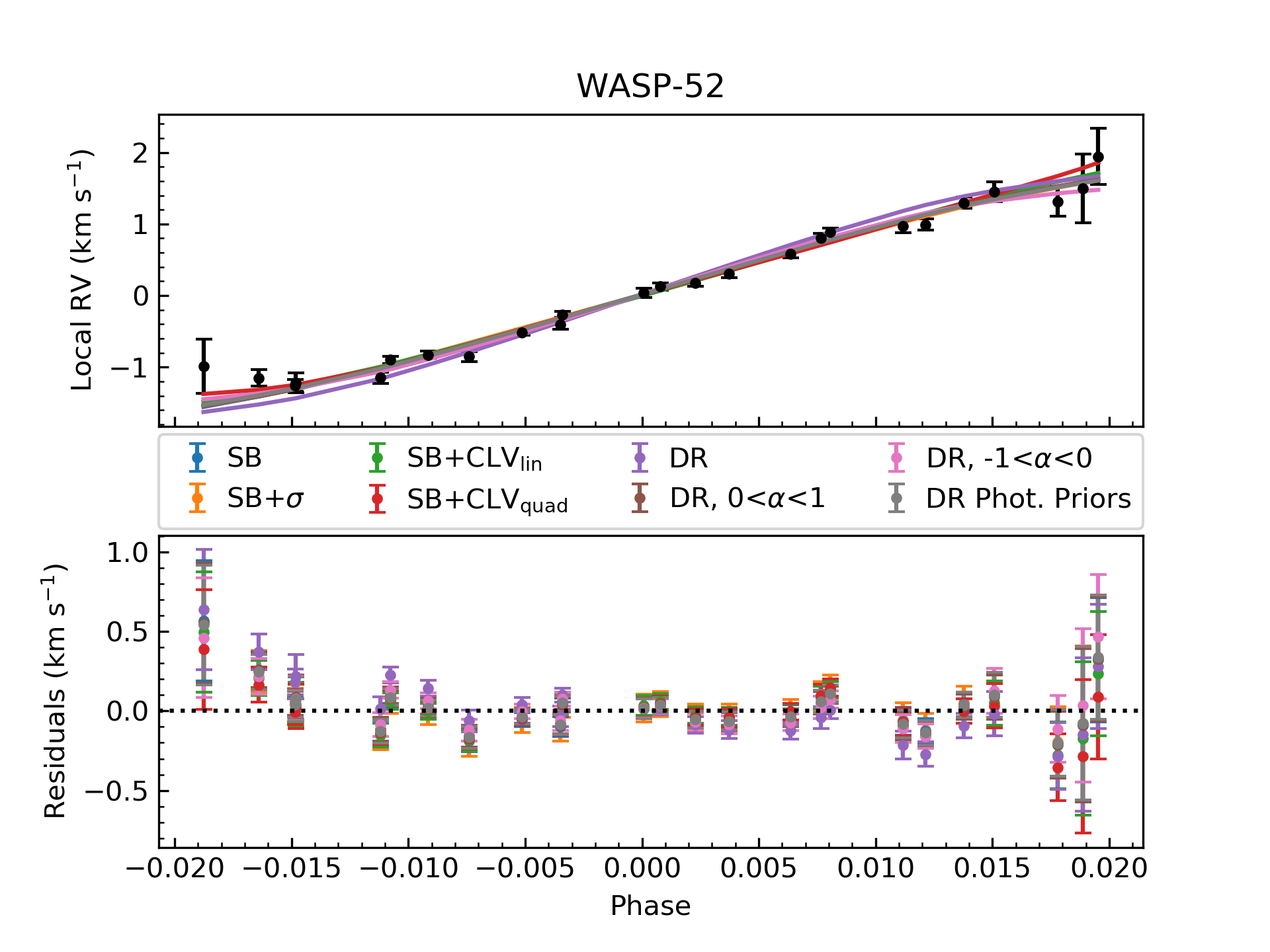}
\caption[]{Top: local RVs as a function of phase (in black), over-plotted with the various model fits for the velocity contributions behind WASP-52~b along the transit chord. Bottom: residuals between the measured local RVs and the aforementioned model fits, with a dotted line at 0~km~s$^{-1}$ to guide the eye.
} 
\label{fig:wasp52_loc_rv}
\end{figure}
\end{center}

The corresponding local RVs for WASP-52, across the transit chord, are displayed in Figure~\ref{fig:wasp52_loc_rv}. If WASP-52~b had occulted a magnetically active region on the host star, then we would expect to see a net redshift relative to the surrounding local RVs, due to the suppression of  convection in said region. Similar to the transit light curve analysis, there is no strong evidence for any spot or plage crossings during these observations; that is to say, it is unlikely that there were any discernible magnetically active regions along the transit chord during these times.  

In Figure~\ref{fig:wasp52_loc_rv}, we also plot the best fit models for the various SB, DR, and CLV scenarios. The respective coefficients for the various fits, their Bayesian Information Criterion (BIC), and reduced chi squared statistic ($\chi^2_r$) are provided in Table~\ref{tab:best_fit_wasp52}. It is important to note that the residuals to the various best fit models under the SB rotation are non-Gaussian, e.g. failing the Shapiro, D'Agostino's K$^2$, and Anderson-Darling tests for normality \citep[see][for further references]{2020SciPy-NMeth}; this is largely due to the points closest to the stellar limb. Hence, an inspection of the BIC must be taken with caution. 

\begin{table*}[t!]
\footnotesize
\caption[]{Best fit results for the models of the local RVs of WASP-52}
\begin{center}
\begin{tabular}{c|c|c|c|c|c|c|c|c|c||c}
    \hline
    \hline

Model~\tablefootmark{a} & $v_{eq}$ (km~s$^{-1}$) &  $i{\star}$ ($\degree$) ~\tablefootmark{b}& $\alpha$ & $\lambda$ ($\degree$)  & c$_1$ (km~s$^{-1}$) & c$_2$ (km~s$^{-1}$) & $\sigma$ (km~s$^{-1}$)  & BIC & $\chi^2_r$ & $\psi$ ($\degree$)\\
    \hline  

SB & 2.06$\pm 0.04^{\rm{b}}$ & 90\tablefootmark{c} & 0$^{\rm{b}}$ & 0.64$\pm 0.62$ & -- & -- & -- & 48.2 & 1.8 & --$^{\rm{b}}$ \\

SB + $\sigma$ & 2.04$\pm 0.05^{\rm{b}}$ & 90\tablefootmark{c} & 0$^{\rm{b}}$ & 0.60$^{+0.93}_{-0.94}$ & -- & -- & 0.06$^{+0.03}_{-0.02}$ & 33.5 & 1.1 & --$^{\rm{b}}$ \\

SB + CLV$_{\rm{lin}}$ & 2.06$\pm 0.04^{\rm{b}}$ & 90\tablefootmark{c} & 0$^{\rm{b}}$ & 0.38$\pm 0.64$ & -0.2$^{+0.13}_{-0.14}$ & -- & -- & 49.2 & 1.8 & --$^{\rm{b}}$ \\

SB + CLV$_{\rm{quad}}$ & 2.07$\pm 0.04^{\rm{b}}$ & 90\tablefootmark{c} & 0$^{\rm{b}}$ & 2.94$^{+2.21}_{-2.20}$ & -1.98$^{+1.49}_{-1.48}$ & 1.41$^{+1.18}_{-1.17}$ & -- & 51.0 & 1.8 & --$^{\rm{b}}$ \\

DR & 1.99$^{+0.60}_{-0.43}$ & 115.93$^{+18.96}_{-47.65}$  & -0.50$^{+0.97}_{-0.33}$  & 0.59$^{+0.64}_{-0.66}$ & -- & -- & -- & 104.6 & 4.4 & 33.53$^{+16.73}_{-16.27}$ \\

DR (0$\leq \alpha \leq$1)  & 2.51$^{+0.18}_{-0.14}$ & 64.58$^{+11.24}_{11.71}$ & 0.71$^{+0.21}_{-0.40}$ & 0.66 $\pm$ 0.57 & -- & -- & -- & 52.0 & 1.9 & 20.92$^{+11.75}_{-10.6}$ \\

DR (-1$\leq \alpha \leq$0)  & 1.82$^{+0.62}_{-0.30}$ & 122.05$^{+16.87}_{-17.26}$ & -0.58$^{+0.26}_{-0.27}$& 0.56$^{+0.67}_{-0.65}$ & -- & -- & --& 49.4 & 1.8 & 37.36$^{+16.44}_{-16.23}$ \\

DR (phot.~prior) & 2.47$^{+0.08}_{-0.10}$ & 67.83$^{+7.83}_{-7.62}$ & 0.73$^{+0.20}_{-0.32}$& 0.66$\pm 0.56$ & -- & -- & --& 50.8 & 1.8 & 17.39$^{+7.61}_{-7.71}$ \\

    \hline
  \end{tabular}
\end{center}
\tablefoot{ \tablefoottext{a}{Assuming uninformative, uniform priors, unless otherwise indicated.} \tablefoottext{b}{$i_{\star}$ is constrained to 0-180$\degree$, and values $>90\degree$ indicate the star’s rotation axis is pointing away from the LOS.} \tablefoottext{c}{Fixed under the assumption of rigid body rotation; we note this means the value in the $v_{eq}$ column for this row corresponds to $v_{eq} \sin i_{\star}$ and that we are unable to determine the 3D obliquity, $\psi$.} }
\label{tab:best_fit_wasp52} 
\end{table*}

\vspace{-20pt}

The non-Gaussian residuals from the pure SB model prompted us to inspect whether our uncertainties might be under-estimated; to test this we explored the best fit of SB rotation and a white noise term, $\sigma$. Introducing an additional white noise term will, by design, push our $\chi^2_r$ towards 1. Nonetheless, the improvement in the BIC suggests that there is evidence that our local RV uncertainties might be under-estimated by as much as 60~m~s$^{-1}$ on average. However, the residuals of the SB+$\sigma$ fit still fail normality tests.

The non-normality of the residuals in these two tests indicate it is possible there may be unaccounted for correlated noise present in the data, e.g. from stellar surface variability. Hence, we also fit for an additional contribution from CLV (i.e. SB+CLV); we test both a linear and quadratic form for the CLV (CLV$_{\rm{lin}}$ and CLV$_{\rm{quad}}$, respectively). In both cases the CLV polynomial coefficients are consistent with zero within 2$\sigma$ (though they are highly correlated with $\lambda$ for the quadratic case). The $v \sin i_{\star}$ and $\lambda$ also remain consistent with those from the pure SB fit and all three scenarios provide similar goodness of it, meaning the extra free parameters in the CLV models cannot be justified.

We further explore whether any constraints can be placed on the potential level of DR on WASP-52, although given the projected alignment of the system, this is likely to be difficult to constrain. Implementing the uninformative, uniform priors from above, we find a DR fit results in a split distribution for $\alpha$, $i_{\star}$, and $v_{eq}$ (though the median value for the latter remains close to the $v \sin i_{\star}$ derived under the SB assumption). The projected obliquity appears robust, with a Gaussian distribution and median value close to zero, as found in the SB fits. We note that the $\alpha$ distributions are clustered towards the extrema of the uniform prior, and are much larger than other DR determinations in the literature, which indicates the fits may not be physical. The most probable explanation for these split distributions is that as $\alpha$ nears 0, the $\sin (\theta_{lat}$) term drops out, which leaves only the $\sin (i_{\star})$ latitudinal dependency, and $\sin (i_{\star}) = \sin(\pi - i_{\star})$. In this instance, we see resulting $i_{\star}$ distributions centred around $\sim 60{\degree}$ and $\sim 120{\degree}$ and a divergence at $\alpha=0$. 

Hence, such bi-modal distributions in $\alpha$ and $i_{\star}$ likely indicate that the spectroscopic transits alone are not precise enough to differentiate between solar and anti-solar rotation and/or a star pointing towards or away from us. As such, we explore fitting for a solar-like and anti-solar DR rotation separately (i.e $\alpha \ge 0$ and $\alpha \le 0$, respectively); both of these instances result in a similar, albeit slightly worse, quality of fit compared to the SB case(s). In each case, $| \alpha |$ still diverges towards 1,  and we are unable to completely break degeneracy between $v_{eq}$ and $\sin i_{\star}$. As a result, we do not assign a physical interpretation of the $\alpha$ derived from these fits. 

We note that following \cite{chaplin19}, we would expect the p-modes for WASP-52 to be averaged out within the given exposure times. Nonetheless, further binning the data can increase the S/N. Since p-modes were not expected to be an issue, and we found no evidence for spot occultations, we opted to bin the data in phase to preserve spatial resolution (in bins of $\sim$0.004). However, we found the results here were consistent with the unbinned data. 

In the case of WASP-52, we can use the long-term photometric monitoring from NGTS in an attempt to further constrain the potential DR. In Section~\ref{sec:LC}, spot modelling with \emph{SOAP} indicated $i_{\star} = 77.1 \pm 15.6{\degree}$ and $P_{rot} = 17.69 \pm 0.12$~d. In turn, we use these to add Gaussian priors on $i_{\star}$ and $v_{eq}$ in a further DR fit of the local RVs. Although $v_{eq}$, $\lambda$ and $i_{\star}$ are well constrained, we are still unable to break the degeneracy between $v_{eq}$ and $\alpha$, with the preferred $\alpha$ approaching the $\alpha = 1$ boundary (with the result largely consistent with the $0 \le \alpha \le 1$ prior). Hence, we do not trust the rotational shear returned from this DR model fit. Unfortunately, even with the slight inclination of the stellar spin axis, the small projected alignment means the planet likely occults only 4${\degree}$ of latitude (half that of the HD~189733~b system); this fact combined with the slow rotation means the absolute change in rotational velocity from any rotational shear along the transit chord is very small. 

\subsection{HAT-P-30}
\begin{figure}[t!]
\includegraphics[trim=0.2cm 9.8cm 0.cm 0.cm, clip, scale=0.65]{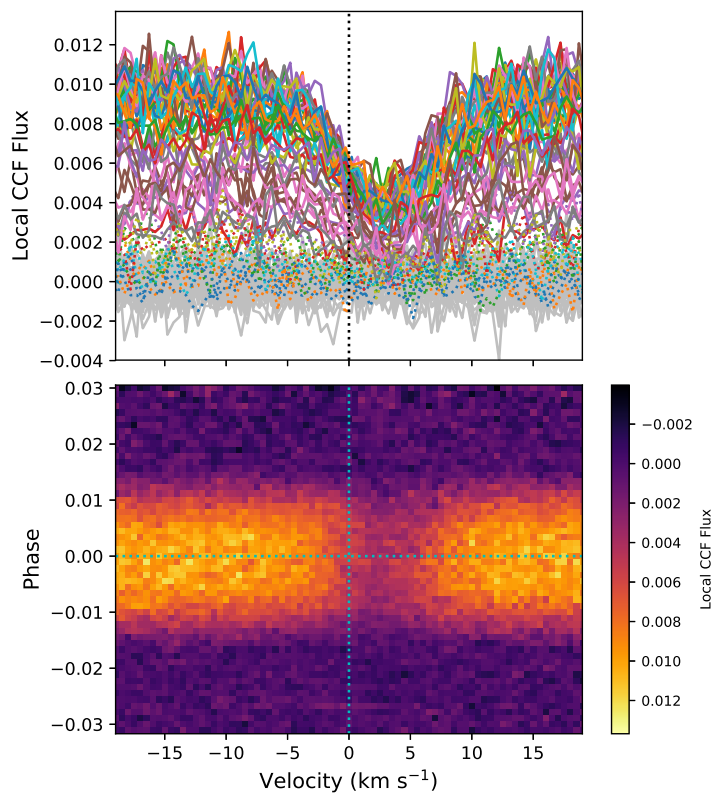}
\includegraphics[trim=0.2cm 0.cm 0.cm 8.7cm, clip, scale=0.65]{HAT-P-30_loc_cff_phasevelflux.eps}
\caption[]{Local stellar CCFs, behind the transiting planet (i.e. CCF$_{\rm{out}}$ - CCF$_{\rm{in}}$), for HAT-P-30. Top: plotted in various colours in the stellar rest frame (indicated by a dotted horizontal line); the two in-transit local CCFs that fail the signal-to-noise test are shown as dashed lines. Individual out-of-transit CCFs subtracted from the master CCF$_{\rm{out}}$ are displayed in grey. Bottom: plotted as a function of phase and colour-coded by the flux; a further horizontal dashed line indicates zero phase.} 
\label{fig:hatp30_loc_ccf}
\end{figure}

The local CCFs behind the transiting planet for HAT-P-30 are shown in Figure~\ref{fig:hatp30_loc_ccf}; the residuals from a subtraction between the master-out and individual out-of-transit CCFs are shown in grey in the top plot and in dark purple in the bottom plot. The local CCFs that fail the quality test are shown as dotted lines in the top of  Figure~\ref{fig:hatp30_loc_ccf}. Of the 61 in-transit local CCFs from the two transits, 48 pass the quality test. Although HAT-P-30 is brighter than WASP-52, the very high impact factor for this system ($b \approx 0.86$) means its planet spends less time fully on the stellar disc and occults regions with much heavier limb darkening, both of which decrease the signal-to-noise of the resulting local CCFs.

\begin{center}
\begin{figure}[t!]
\centering
\includegraphics[trim=0.4cm 0.3cm 1.5cm 0.9cm, clip, scale=0.63]{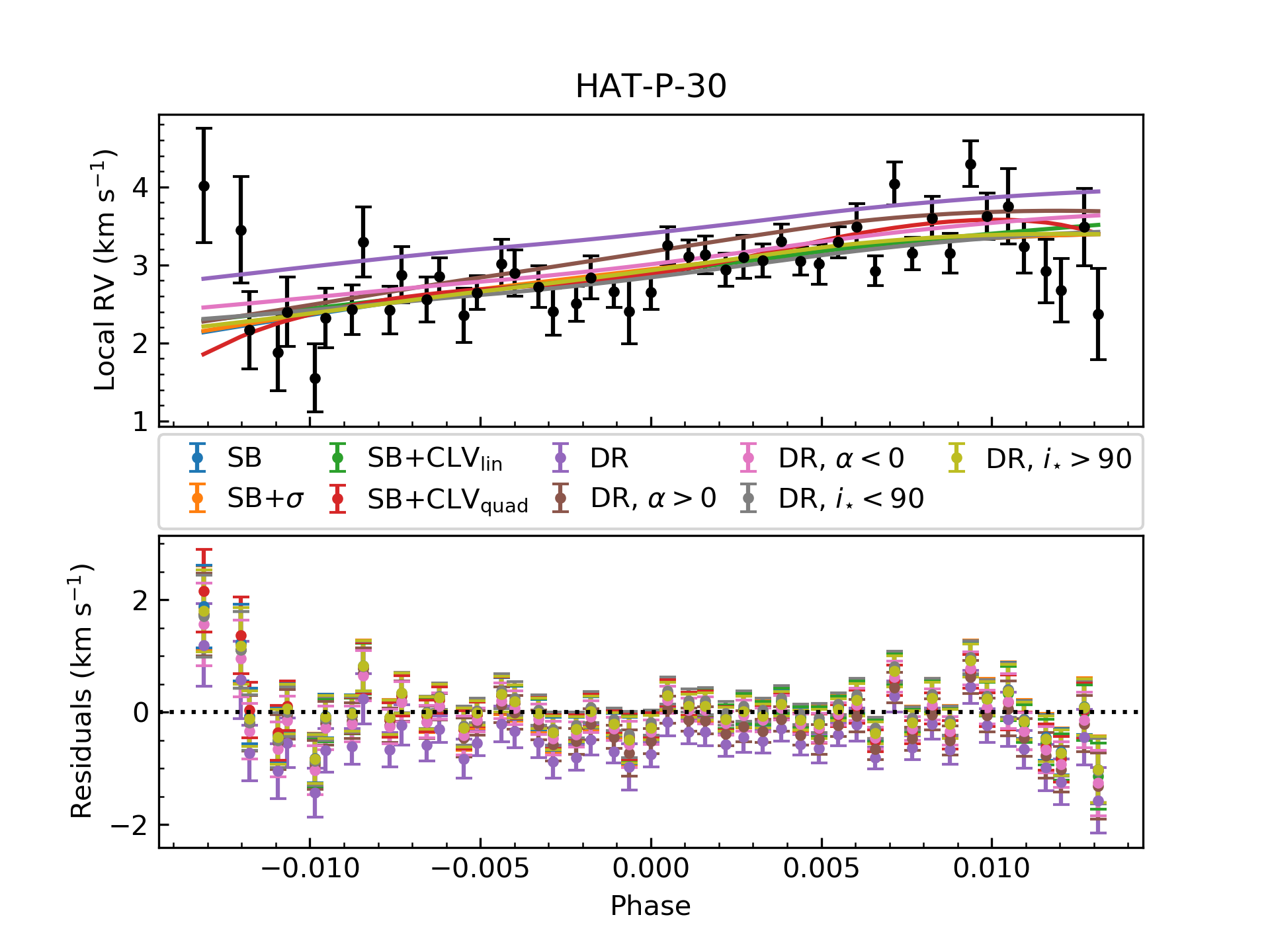}
\caption[]{Top: local RVs as a function of phase (in black), over-plotted with the various model fits for the velocity contributions behind HAT-P-30~b along the transit chord. Bottom: residuals between the measured local RVs and the aforementioned model fits, with a dotted line at 0~km~s$^{-1}$ to guide the eye. We note the DR fit without additional priors is a poor fit due to the bi-modal distributions in the fitted parameters, see main text and Appendix~\ref{apn:mcmc_hatp30} for details. 
} 
\label{fig:hatp30_loc_rv}
\end{figure}
\end{center}

The projected misalignment of this system is clearly indicated in Figure~\ref{fig:hatp30_loc_ccf}, as the centroid of the local CCFs all lie to one side of the stellar rest frame; this is further evidenced by the corresponding local RVs plotted in Figure~\ref{fig:hatp30_loc_rv}. At this stage in the initial analysis we noted a significant offset between the two transits, which was remedied by utilising the sky-subtracted observations for the second transit. 

In Figure~\ref{fig:hatp30_loc_rv}, we also plot the best fit models for the various SB, DR, and CLV scenarios. The respective coefficients for the various fits, their BIC, and reduced chi squared statistic ($\chi^2_r$) are provided in Table~\ref{tab:best_fit_hatp30}. Similar to the WASP-52 analysis, the residuals to the various best fit models under the SB rotation are non-Gaussian, e.g. failing the Shapiro, D'Agostino's K$^2$, and Anderson-Darling tests for normality, and an inspection of the BIC must be taken with caution. 

Despite the slightly higher $v \sin i_{\star}$, larger projected obliquity, and increased brightness compared to WASP-52, the best-fit model is similarly the SB+$\sigma$, indicating here that the uncertainties may potentially be underestimated by up to $\sim 130$~m~s$^{-1}$. There are likely a combination of additional factors at play here. For example, the $R_p/R_{\star}$ is smaller for WASP-52~b, and the impact factor of HAT-P-30~b is much higher, meaning the planet is nearly always grazing and always occulting regions of high limb darkening. The higher effective temperature of HAT-P-30 also means this star will have larger p-mode and granulation variability, contributing to higher red noise in the local RVs; this is evident in Figures~\ref{fig:wasp52_loc_rv} and \ref{fig:hatp30_loc_rv} where the median RV uncertainties are $\sim 70$ and $\sim 290$~m~s$^{-1}$, respectively. 

For HAT-P-30, the posteriors are well-defined in the SB and SB+$\sigma$ model fits (e.g. see Figures~\ref{fig:hatp30_sb_mcmc} and \ref{fig:hatp30_sb_jitter_mcmc}), but there is a strong correlation between the $v \sin i_{\star}$ and the $c_1$ coefficient for the SB+CLV$_{\rm{lin}}$ fit (see Figure~\ref{fig:wasp-52_sb_cb_lin_mcmc}). Interestingly, when investigating the more flexible quadratic CLV contribution, the CLV coefficients are both linearly correlated with $v \sin i_{\star}$, but there is a split in this distribution resulting in a V-shaped correlation (see Figure~\ref{fig:hatp30_sb_cb_quad_mcmc}). The $\lambda$ and $v \sin i_{\star}$ relationship also goes from well-defined in the linear CLV model to a banana shape in the quadratic CLV model, where the centroid of the banana occurs near $\lambda = 0{\degree}$. The quadratic CLV coefficients are also now strongly correlated with $\lambda$, which is no longer reliably recovered ($\lambda = 9.94^{+45}_{-53}{\degree}$). The behaviour of the SB and CLV fits indicates that at high impact factors, the stellar rotation and centre-to-limb convective contribution may become degenerate if the observations are of insufficient RV precision compared to their relative contributions. 

The initial DR fit reveals a binary distribution in $i_{\star}$, similar to what we saw for WASP-52, but the $\alpha$ distribution is relatively flat this time (as opposed to the binary distribution seen in WASP-52). As such, we explore both solar/anti-solar priors and priors to enforce $i_{\star}$ to either point towards or away from the line-of-sight. In all cases, the projected obliquity and equatorial velocity remain in agreement within 1$\sigma$. When $i_{\star}$ is forced to point towards the observer, there is broad agreement with the anti-solar DR fit, with $i_{\star} \approx$~50-60${\degree}$. Similarly, the solar DR fit is in broad agreement when forcing $i_{\star}$ to point away, with $i_{\star} \approx$~120-140${\degree}$. On the other hand, the solar and anti-solar $\alpha$ distributions are relatively flat, while the $i_{\star}$ priors resulted in $\alpha$ distributions that favoured the extrema of the uniform priors. Nonetheless, in all cases $\alpha = 0$ cannot be disregarded and the BIC favours the solid body solutions. Moreover, it is clear from Figure~\ref{fig:hatp30_loc_rv}, that the difference in the various model fits is small relative to the scatter in the data.

As HAT-P-30 was the more ideal target to search for DR and CLV (owing to its larger obliquity and higher effective temperature and lower magnetic activity), for this target we also tested the DR+CLV fits for various CLV polynomials up to cubic. This was partially inspired by \cite{doyle22}, where the authors found a potentially large CLV on the hot F star WASP-166 could dominate over the DR effects; in this case, a DR fit could only be preferred if both DR and a cubic CLV were fit simultaneously. However, in each case here the CLV coefficients were still always consistent with zero, the $\alpha$ distribution was relatively flat and the $i_{\star}$ distribution split around 90${\degree}$, as before. It is interesting to note that $\lambda$ was only recovered around 70${\degree}$ when considering the linear CLV; for the higher orders the preferred obliquity was near 40-50${\degree}$. For completeness, we also tested a SB plus cubic CLV fit; as expected we found the CLV coefficients consistent with zero, but the $\lambda$ distribution was then centred closer to 60${\degree}$ with a large tail towards lower obliquities, effectively cutting off some of the CLV-$\lambda$ degeneracy seen in the SB+CLV$_{quad}$ fit. In all cases, we could not justify the extra degrees of freedom required by these models. 

\vspace{-30pt}

\begin{table*}[t!]
\footnotesize
\caption[]{Best fit results for the models of the local RVs of HAT-P-30, assuming uninformative, uniform priors}
\begin{center}
\begin{tabular}{c|c|c|c|c|c|c|c|c|c||c}
    \hline
    \hline

Model & $v_{eq}$ (km~s$^{-1}$) &  $i{\star}$ ($\degree$) ~\tablefootmark{a}& $\alpha$ & $\lambda$ ($\degree$)  & c$_1$ (km~s$^{-1}$) & c$_2$ (km~s$^{-1}$) & $\sigma$ (km~s$^{-1}$)  & BIC & $\chi^2_r$ & $\psi$ ($\degree$)\\
    \hline  

SB & ${3.62^{+0.07}_{-0.06}}^{\rm{b}}$ & 90\tablefootmark{b} & 0$^{\rm{b}}$ & 70.19$^{+2.54}_{-2.46}$ & -- & -- & -- & 74.8 & 1.5 & --$^{\rm{b}}$ \\

SB + $\sigma$ & 3.63$\pm0.07^{\rm{b}}$ & 90\tablefootmark{b} & 0$^{\rm{b}}$ & 70.49$^{+2.86}_{-2.77}$ & -- & -- & 0.13$ \pm 0.08$ & 69.3 & 1.3 & --$^{\rm{b}}$ \\

SB + CLV$_{\rm{lin}}$ & 3.41$\pm 0.23^{\rm{b}}$ & 90\tablefootmark{b} & 0$^{\rm{b}}$ & 69.45$^{+2.83}_{-2.81}$ & -0.69$^{+0.71}_{-0.69}$ & -- & -- & 77.8 & 1.5 & --$^{\rm{b}}$ \\

SB + CLV$_{\rm{quad}}$ & ${1.53^{+0.87}_{-0.35}}^{\rm{b}}$ & 90\tablefootmark{b} & 0$^{\rm{b}}$ & 9.94$^{+45.15}_{-53.02}$ & 16.03$^{+7.42}_{-8.30}$ & -19.85$^{+9.80}_{-8.72}$ & -- & 85.2 & 1.6 & --$^{\rm{b}}$ \\

DR & 4.51$^{+1.47}_{-0.68}$ & 66.17$^{+68.88}_{-23.20}$ &  -0.28$^{+0.67}_{-0.53}$ &  73.19$^{+8.41}_{-7.01}$ & -- & -- & -- & 238.7 & 5.1 & 74.27$^{+13.79}_{-7.49}$ \\

DR (0$\leq \alpha \leq$1) & 5.08$^{+2.28}_{-1.30}$ & 126.94$^{+20.67}_{-60.44}$ & 0.30$^{+0.41}_{-0.23}$ & 75.92$^{+8.89}_{-5.70}$ & -- & -- & -- & 123.3 & 2.5 & 84.56$^{+8.62}_{-13.93}$ \\

DR (-1$\leq \alpha \leq$0)  & 4.40$^{+1.17}_{-0.61}$ & 57.01$^{+50.82}_{-16.00}$ & -0.49$^{+0.37}_{-0.36}$ & 70.47$^{+6.63}_{-6.27}$ & -- & -- & -- & 90.5 & 1.7 & 72.41$^{+6.24}_{-6.99}$ \\

DR (0$\leq i_{\star} \leq$90)  & 4.38$^{+1.01}_{-0.55}$ & 51.94$^{+19.53}_{-13.02}$ & -0.55$^{+0.45}_{-0.33}$ & 71.60$^{+6.58}_{-5.63}$ & -- & -- & -- & 82.2 & 1.5 & 71.6$^{+5.06}_{-5.83}$ \\

DR (90$\leq i_{\star} \leq$180)  & 5.43$^{+2.26}_{-1.66}$ & 135.97$^{+13.60}_{-30.46}$ & 0.33$^{+0.42}_{-0.40}$ & 77.66$^{+7.93}_{-9.28}$ & -- & -- & -- & 80.8 & 1.5 & 87.89$^{+5.74}_{14.32}$ \\

    \hline
  \end{tabular}
\end{center}
\tablefoot{ \tablefoottext{a}{$i_{\star}$ is constrained to 0-180$\degree$, and values $>90$$\degree$ indicate the star’s rotation axis is pointing away from the LOS.} \tablefoottext{b}{Fixed under the assumption of rigid body rotation; we note this means the value in the $v_{eq}$ column for this row corresponds to $v_{eq} \sin i_{\star}$ and that we are unable to determine the 3D obliquity, $\psi$.} 
}
\label{tab:best_fit_hatp30}
\end{table*}

As a final effort, we also tried binning the data to reduce the impact of p-modes. Following the methodology in \cite{chaplin19}, we expect the p-modes to have a negligible effect with exposure times near 8~minutes. To very roughly account for uncertainties discussed in \cite{chaplin19}, we opted to bin in approximately 20~minute intervals. We repeated the analysis above on the binned data, but found that the parameters were consistent within 1$\sigma$. The null effect of binning here may indicate that the dominate issues are not due to p-modes, but rather the geometry of the system (e.g. the high impact factor) and/or precision of the observations which still limit this system even after binning, perhaps due to red noise from the granulation.  

\section{Simulations}
\label{sec:sim}
Given the relatively low effective temperature of WASP-52, combined with its slow rotation and the low projected obliquity, we did not expect to detect DR or CLV for this target (we remind the reader the aim was to explore spot occultations, but there was insufficient evidence of occultations for the transits herein). However, given both the higher effective temperature and large projected misalignment of HAT-P-30, our aim was to detect both DR and CLV for this target. As shown in Section~\ref{sec:RM}, we did not find sufficient evidence for either phenomena in either of our targets. Here, we briefly simulate the HAT-P-30 system to explore further under what scenarios we should be able to detect DR and CLV if present at a solar-like level. In particular, we explore a variety of projected obliquities, stellar inclinations, stellar rotation rates, and impact factors.

We follow the procedure in \cite{Roguet-Kern22} to simulate local RVs with appropriate uncertainties and test DR/CLV retrievals; where not stated the parameters are fixed to those in Table~\ref{tab:hatp30_param} for the HAT-P-30 system. In our first simulations, we explore the conditions where we might expect to confidently detect DR if present at the solar level (i.e. a differential shear = 0.2), neglecting any CLV contributions. We set $v \sin i_{\star}$ = 3.6 km~s$^{-1}$ and explore the whole $\lambda$ parameter space (-180 to $+$180${\degree}$, in 4${\degree}$ steps). We vary $i_{\star}$ between 0-180${\degree}$ in 2${\degree}$ steps, and $v_{eq}$ is then derived given the fixed $v \sin i_{\star}$ value above.  We use the time sampling of the observations herein, and the local RV uncertainties are set by Equation~1 in \cite{Roguet-Kern22}: 

  \begin{equation}
   \label{eqn:scale}
       RV_{err}(\mu) = \sqrt{\frac{1}{\rm{gain}}\cdot\left(\frac{1}{R_*}\right)^2\cdot10^{\frac{V}{2.5}}\cdot\frac{1}{t_{exp}}}\cdot\frac{LC(\mu)}{1 - LC(\mu)}\cdot C,
   \end{equation}
 
 where the gain was set to 6 for ESPRESSO\footnote{The gain is set to 1 for HARPS, under which the original expression was derived.}, $R_*$ is the stellar radius, $t_{exp}$ is the exposure time in seconds, and LC is the light curve (which is also a function of centre-to-limb angle, $\mu$). In \cite{Roguet-Kern22}, $C$ is a constant scaled to the local uncertainties of HD~189733 to account for other instrumental setup and environmental parameters. At the time, this constant was found to accurately predict the local RV uncertainties from a RRM analysis for various star-planet systems. This is indeed the case for WASP-52, but here we find using the same $C$ systematically under predicts the uncertainties for HAT-P-30. All three systems have a similarly slow rotation, but HAT-P-30 is much hotter so the convective broadening of the local CCF is much larger and may lead to a decease in the overall RV precision. Advancing Equation~\ref{eqn:scale} is beyond the scope of this paper and will be the subject of future work; herein we scale $C$ to the empirically derived uncertainties measured in Section~\ref{sec:RM}. We opt to use the predicted uncertainties, rather than the empirical, as they represent the most ideal observing conditions and the most ideal scenario of pure white noise, thereby setting an lower limit. Following \cite{Roguet-Kern22}, we use the difference in BIC assess goodness of fit. To account for different realisations of white noise, we repeat the simulation and fit three times and take the average. 

As expected, the uncertainties are too large to justify the more complicated DR model over SB regardless of any potential stellar inclination or projected obliquity. The most likely limiting factors for a DR detection are the low projected stellar rotation rate and the high impact factor. Hence, we also simulate scenarios with either a $v \sin i_{\star}$ = 5~km~s$^{-1}$ and 10~km~s$^{-1}$, or the measured rotation $v \sin i_{\star}$ = 3.6~km~s$^{-1}$ and an impact factor of b = 0.2, 0.4, or 0.6. 

We find that either a faster rotation or lower impact factor could indeed enable a DR detection. For the $v \sin i_{\star}$ = 10~km~s$^{-1}$ (b $\approx$ 0.87) and b = 0.6-0.4 ($v \sin i_{\star}$ = 3.6~km~s$^{-1}$) cases there are some combinations of $\lambda$ and $i_{\star}$ where a DR detection may be feasible. For the known obliquity near 70${\degree}$, there may be potential for a DR detection if the stellar inclination were near 90${\degree}$ or the star was pole-on. We note that the accuracy of the retrieved differential shear, $\alpha$, is improved for lower impact factors, except at b = 0.2 where the absolute value of the stellar rotation rate along the transit chord (for the known obliquity) is low and thus DR becomes difficult (or impossible) to detect (e.g. shown in \cite{Roguet-Kern22}).  

We also tested these various scenarios under the assumption of the (lower) uncertainty derived when the scalar $C$ was derived from HD~189733. Naturally, this led to a larger DR detection parameter space, where we would have expected to detect DR for HAT-P-30. Hence, if the star was potentially cooler, it may have been possible to detect DR (if present at the solar level) at the real impact factor and stellar rotation rate. 

To explore the potential for a CLV detection for HAT-P-30, we first took a DR+CLV model and added the subsequent white noise (as described above). We explored the same parameter space as before, except we kept $\lambda$ near the known value, exploring between 50-100${\degree}$. We use the CLV from \cite{Roguet-Kern22} based off the net 0~G solar MHD simulations in \cite{cegla18}. We note that for a hot, F dwarf like HAT-P-30, this CLV is likely an under-estimate (e.g. see \citealt{beeck13, doyle22}).

As the CLV is on the same order as the DR, we expect our simulated injeciton/recory to perform similarly to the pure DR case; this was indeed the case. The high impact factor of HAT-P-30~b means that planet track is always at high limb angles (low $\mu$), making it difficult to trace the full centre-to-limb behaviour. However, even for the simulations with $b=0.6$ or $0.4$, we were still unable to prefer a DR+CLV model given the precision of the simulated data. If instead we had a higher precision dataset, e.g. matching that scaled from HD~189733, then we predict a CLV detection if b = 0.4 and the stellar rotation axis was perpendicular or the pole-on. 

We also explored a simplified case, where the star rotated as a solid body (with the rate above) and had a solar-like CLV. In this case, we predict a CLV detection in the entire parameter space explored above, in other words, a SB+CLV fit is preferred over a SB fit. If the star rotated as a solid body, any latitudinal variation in RV could only originate from a CLV contribution. In the DR case, the latitudinal contribution could come from the rotational shearing or the CLV; these extra degrees of freedom require a higher precision to disentangle. Hence, for HAT-P-30 it is likely that either the CLV level is below that of the solar values explored here or the contribution of DR prevents a CLV detection given the precision of the current dataset. We note that since we only simulated one functional form of CLV, we cannot comment on the detectability of a CLV with a different shape. 

\section{Summary and concluding remarks}
\label{sec:conc}
We analysed spectroscopic and photometric transits of WASP-52~b and HAT-P-30~b to examine the respective host stars' stellar surface phenomena. As WASP-52~b has often occulted starspots in the past, our goal was to use the reloaded RM technique to isolate the starlight from such regions and examine the spot properties directly. A secondary goal for this target was to explore potential differential rotation and convective centre-to-limb variations; given the slow rotation of this system, the moderate/low effective temperate and high magnetic activity of the K dwarf host star this was foreseen to be challenging. However, a DR and CLV detection was the primary goal for the hotter F star HAT-P-30, with its projected obliquity previously known to be near 70${\degree}$; in this case the higher effective temperature of the host star (and lower magnetic activity) is expected to lead to a larger (more easily detectable) CLV, while a higher obliquity means the planet is likely to occult more stellar latitudes and therefore more easily enable a DR detection.  

From the photometric transits (using Eulercam and NGTS for both targets, and additional TESS data for HAT-P-30), we have updated the ephemeris for each system and the corresponding transit parameters. For WASP-52, we also performed long-term photometric monitoring with NGTS to further refine the stellar rotation period, which we found to be 17.69~$\pm$~0.12~d (in agreement with the literature). Unfortunately, we find no evidence for spot occultations during the transits observed herein for WASP-52~b.

For both systems, the preferred stellar surface model is solid body rotation plus a white noise term. We note that the p-modes are expected to be averaged out within the individual exposures for WASP-52. For HAT-P-30, we expect the p-modes to be negligible with larger exposure times and examined the data binned on 20~minute intervals, but found no significant changes in the returned model fits. All stellar surface model fits that included a CLV polynomial returned coefficients consistent with zero. When considering a linear CLV contribution, $\lambda$ was unaffected. However, a quadratic CLV fit for HAT-P-30 resulted in a split distribution in $\lambda$, indicating the stellar rotation and CLV may be degenerate at high impact factors if RV precision is insufficient. An attempt to recover DR resulted in bi-modal distributions for both targets. To overcome these degeneracies for WASP-52, we tried utilising Gaussian priors on $i_{\star}$ and $P_{rot}$ (from the photometry) and also restricting the parameter space to either a purely solar-like or anti-solar differential shear. Whilst restricting the parameter space did improve the goodness of fit, it was not sufficient to justify the extra degrees of freedom. For HAT-P-30, we did not have a prior on the stellar rotation, so we further restricted the DR parameter space to explore a solar-like vs anti-solar shear and whether the star was pointing towards us or away; of these, a solar-like shear with the star pointing towards the observer was preferred, but not over solid body rotation. For HAT-P-30, we also tried to fit DR and CLV simultaneously, in case both effects are on the same order, but we were still unable to constrain either effect. 

Due to the somewhat surprising result for HAT-P-30, we simulated various scenarios to try to identify when DR and/or CLV might be detectable, if present at the solar level. Given the precision of our dataset for HAT-P-30, we predict that a DR detection would only be possible if the star was either spinning much faster (e.g. near 10~km~s$^{-1}$) or the impact factor was lower (e.g. between $b=0.4$ to $0.6$). However, at too low of an impact factor (e.g. $b=0.2$), the transit chord covers very low stellar rotation rates, given the high projected obliquity, and thus DR would become challenging or impossible. We postulate that the limiting factors for the precision may be the host star brightness, red noise due to granulation (which is higher for hotter stars), and/or the high impact factor which means the transit chord is heavily limb darkened and the planet is nearly always grazing. If the host star had been more similar to HD~189733, we may have been able to recover a DR+CLV detection if the stellar inclination was either perpendicular or pole-on. Interestingly, if the true stellar rotation could be represented by a solid body, then we predict that we would have expected a SB+CLV detection if the CLV were present at the solar level (or higher). This indicates that either the CLV for this system is below the solar values or possibly that the DR present is preventing a CLV confirmation given the precision of the current dataset.  

Further observations and/or a sophisticated treatment of the red noise present in these observations is required to refine the potential differential rotation and convective centre-to-limb variations present on WASP-52 and HAT-P-30. Future observations of WASP-52 also present a new opportunity to capture starspot occultations and directly probe the spot properties. Moreover, we advocate for a thorough and cautious approach when searching for these subtle stellar signatures as various degeneracies are possible; this is especially true for systems with high impact factors (i.e grazing transits).    

\begin{acknowledgements} 
HMC, NRK, VB, ML and CL, BA acknowledge the financial support of the National Centre for Competence in Research ``PlanetS'' supported by the Swiss National Science Foundation (SNSF). This research has made use of NASA's Astrophysics Data System Bibliographic Services. HMC was supported during this project from both an NCCR PlanetS CHEOPS Fellowship and a UKRI Future Leaders Fellowship (MR/S035214/1). This project has received funding from the European Research Council (ERC) under the European Union's Horizon 2020 research and innovation programme (project {\sc Spice Dune}, grant agreement No 947634). R. A. is a Trottier Postdoctoral Fellow and acknowledges support from the Trottier Family Foundation; this work was supported in part through a grant from FRQNT. The authors would like to thank Dan Foreman-Mackey for useful discussions and the anonymous referee for their constrictive report which help improved the manuscript.
\end{acknowledgements}

\bibliographystyle{aa}
\bibliography{mybib}

\begin{appendix}
\label{appen:wasp52}

\section{Posterior Distributions for WASP-52}
\label{apn:mcmc_wasp52}

\begin{center}
\begin{figure*}[t!]
\centering
\includegraphics[trim=0.cm 0.cm 0.cm 0.cm, clip, scale=0.63]{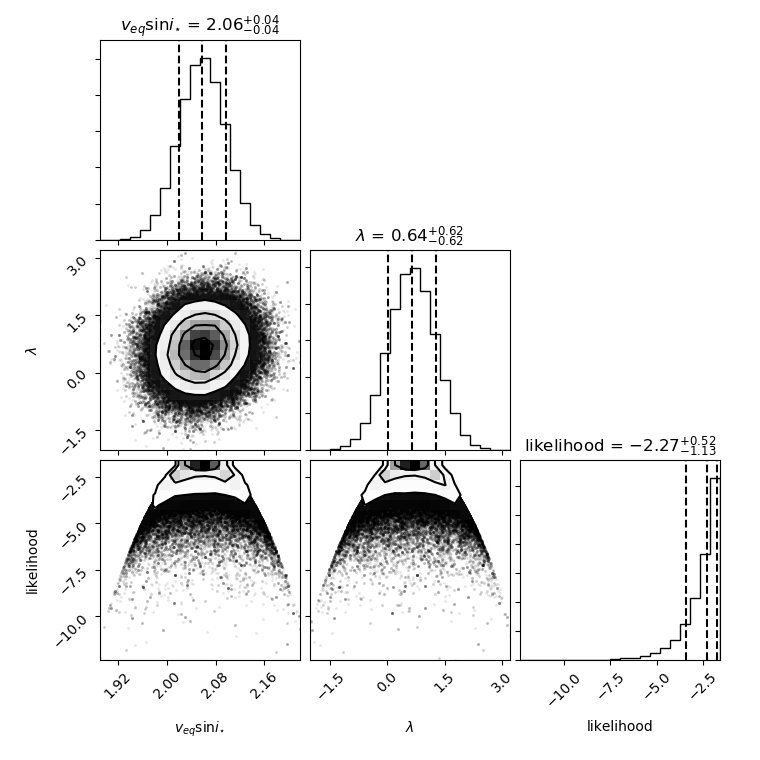}
\caption[]{
WASP-52 corner plot for solid body stellar rotation (SB); displayed are the one and two dimensional projections of the posterior probability distributions and their corresponding log-normal likelihood estimates. Vertical dashed lines correspond to the median values and their respective 1$\sigma$ uncertainties calculated from the 16th, 50th, and 84th percentiles of the samples.   
} 
\label{fig:wasp-52_sb_mcmc}
\end{figure*}
\end{center}

\begin{center}
\begin{figure*}[t!]
\centering
\includegraphics[trim=0.cm 0.cm 0.cm 0.cm, clip, scale=0.63]{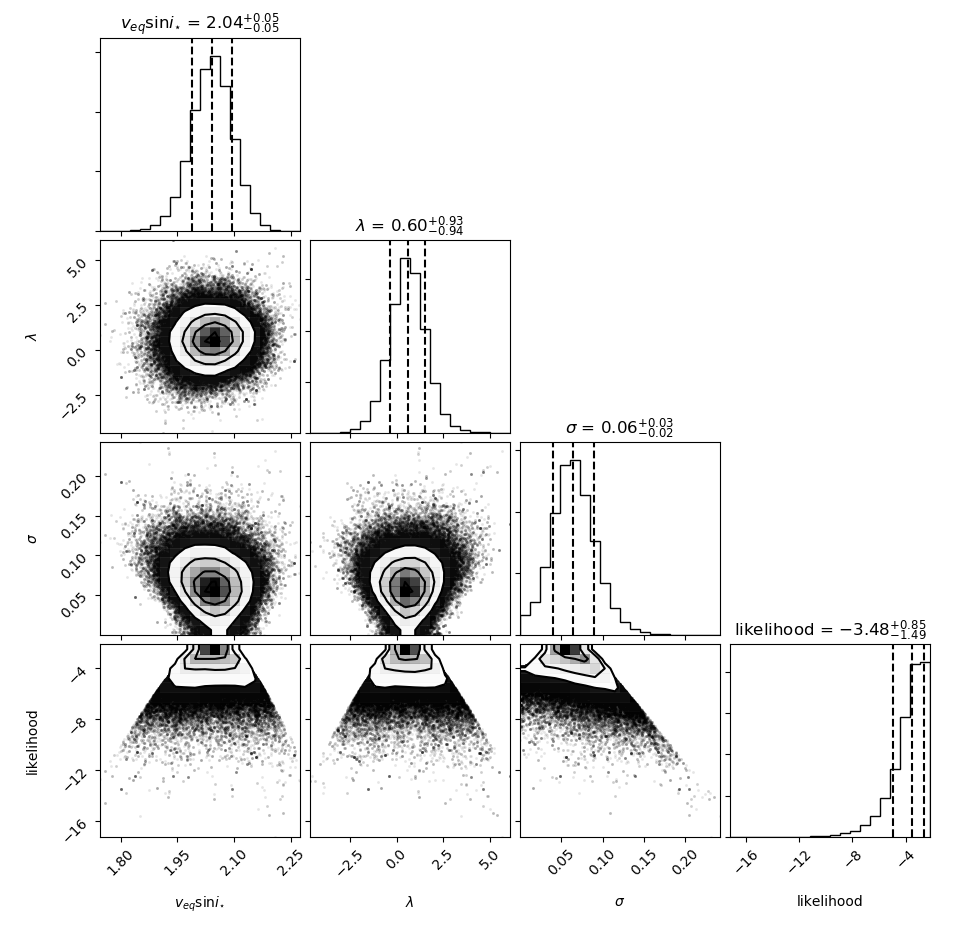}
\caption[]{
WASP-52 corner plot for solid body stellar rotation with an additional white noise term (SB+$\sigma$); displayed are the one and two dimensional projections of the posterior probability distributions and their corresponding log-normal likelihood estimates. Vertical dashed lines correspond to the median values and their respective 1$\sigma$ uncertainties calculated from the 16th, 50th, and 84th percentiles of the samples.   
} 
\label{fig:wasp-52_sb_jitter_mcmc}
\end{figure*}
\end{center}

\begin{center}
\begin{figure*}[t!]
\centering
\includegraphics[trim=0.cm 0.cm 0.cm 0.cm, clip, scale=0.63]{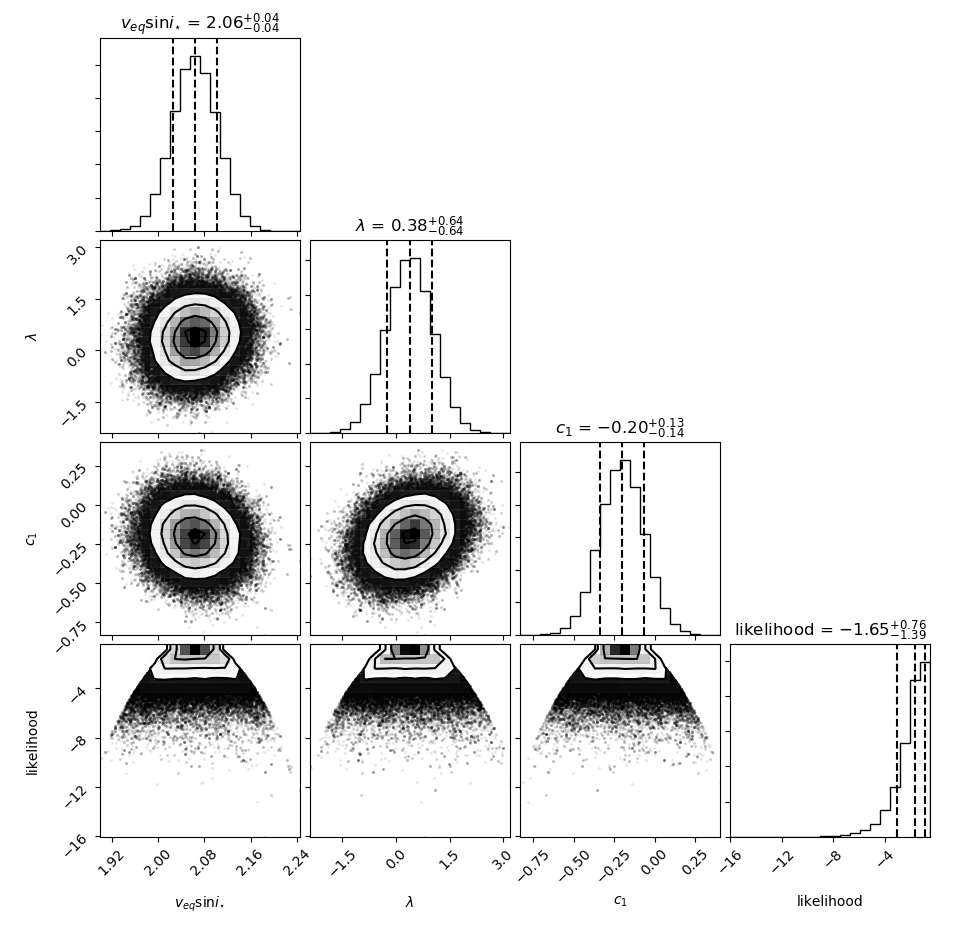}
\caption[]{
WASP-52 corner plot for solid body stellar rotation with a linear centre-to-limb variation due to convection (SB+CLV$_{\rm{lin}}$); displayed are the one and two dimensional projections of the posterior probability distributions and their corresponding log-normal likelihood estimates. Vertical dashed lines correspond to the median values and their respective 1$\sigma$ uncertainties calculated from the 16th, 50th, and 84th percentiles of the samples.   
} 
\label{fig:wasp-52_sb_cb_lin_mcmc}
\end{figure*}
\end{center}

\begin{center}
\begin{figure*}[t!]
\centering
\includegraphics[trim=0.cm 0.cm 0.cm 0.cm, clip, scale=0.63]{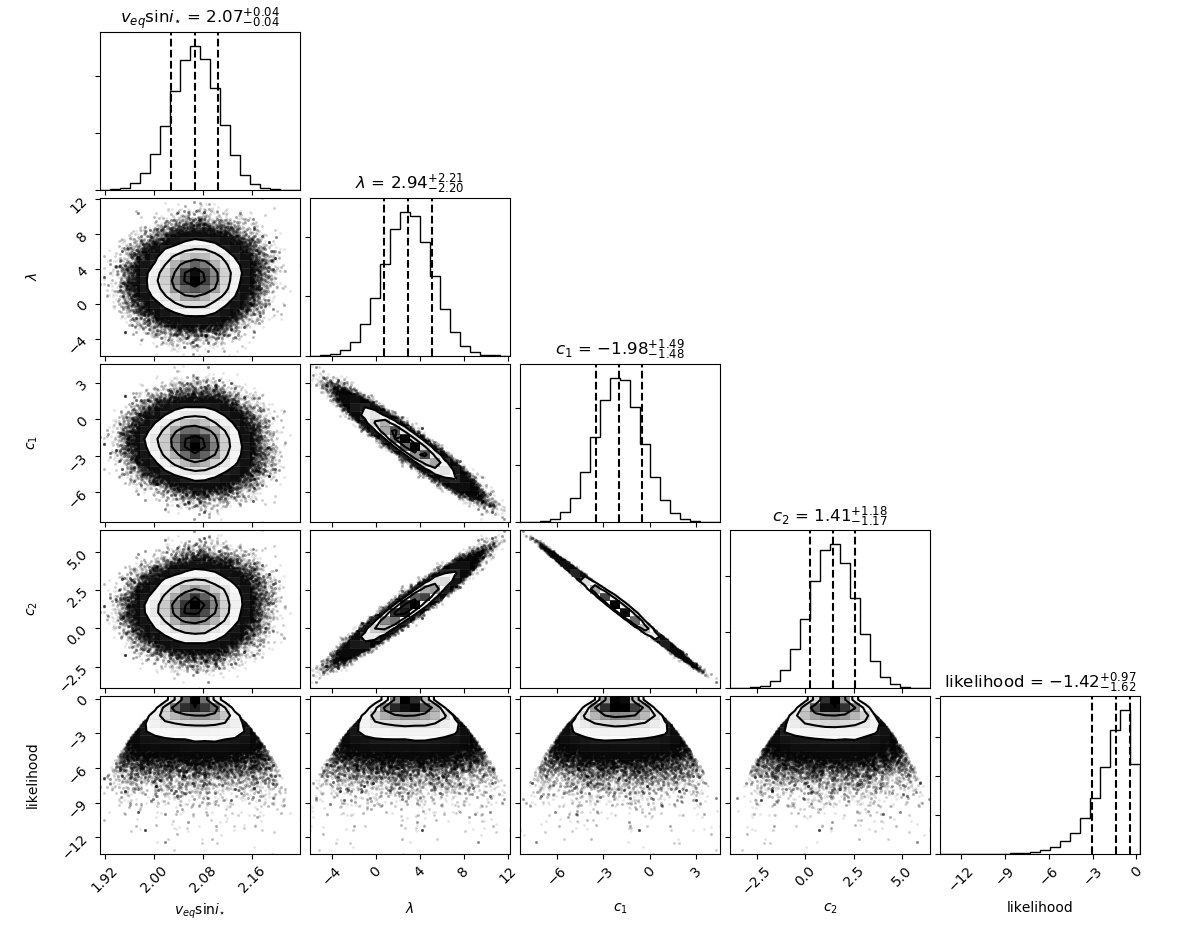}
\caption[]{
WASP-52 corner plot for solid body stellar rotation with a quadratic centre-to-limb variation due to convection (SB+CLV$_{\rm{quad}}$); displayed are the one and two dimensional projections of the posterior probability distributions and their corresponding log-normal likelihood estimates. Vertical dashed lines correspond to the median values and their respective 1$\sigma$ uncertainties calculated from the 16th, 50th, and 84th percentiles of the samples.   
} 
\label{fig:wasp-52_sb_cb_quad_mcmc}
\end{figure*}
\end{center}

\begin{center}
\begin{figure*}[t!]
\centering
\includegraphics[trim=0.cm 0.cm 0.cm 0.cm, clip, scale=0.63]{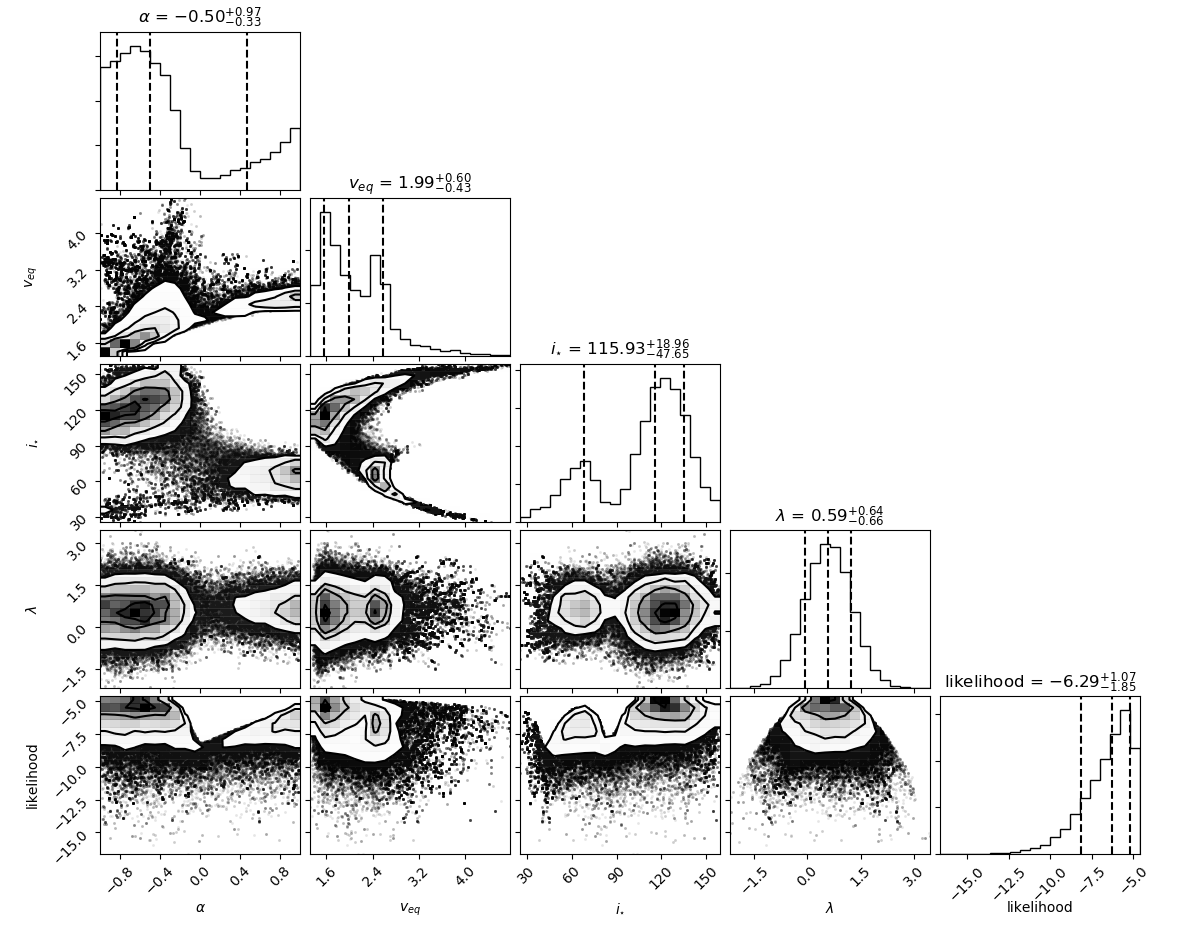}
\caption[]{
WASP-52 corner plot for differential stellar rotation (DR); displayed are the one and two dimensional projections of the posterior probability distributions and their corresponding log-normal likelihood estimates. Vertical dashed lines correspond to the median values and their respective 1$\sigma$ uncertainties calculated from the 16th, 50th, and 84th percentiles of the samples.   
} 
\label{fig:wasp-52_dr_mcmc}
\end{figure*}
\end{center}

\begin{center}
\begin{figure*}[t!]
\centering
\includegraphics[trim=0.cm 0.cm 0.cm 0.cm, clip, scale=0.63]{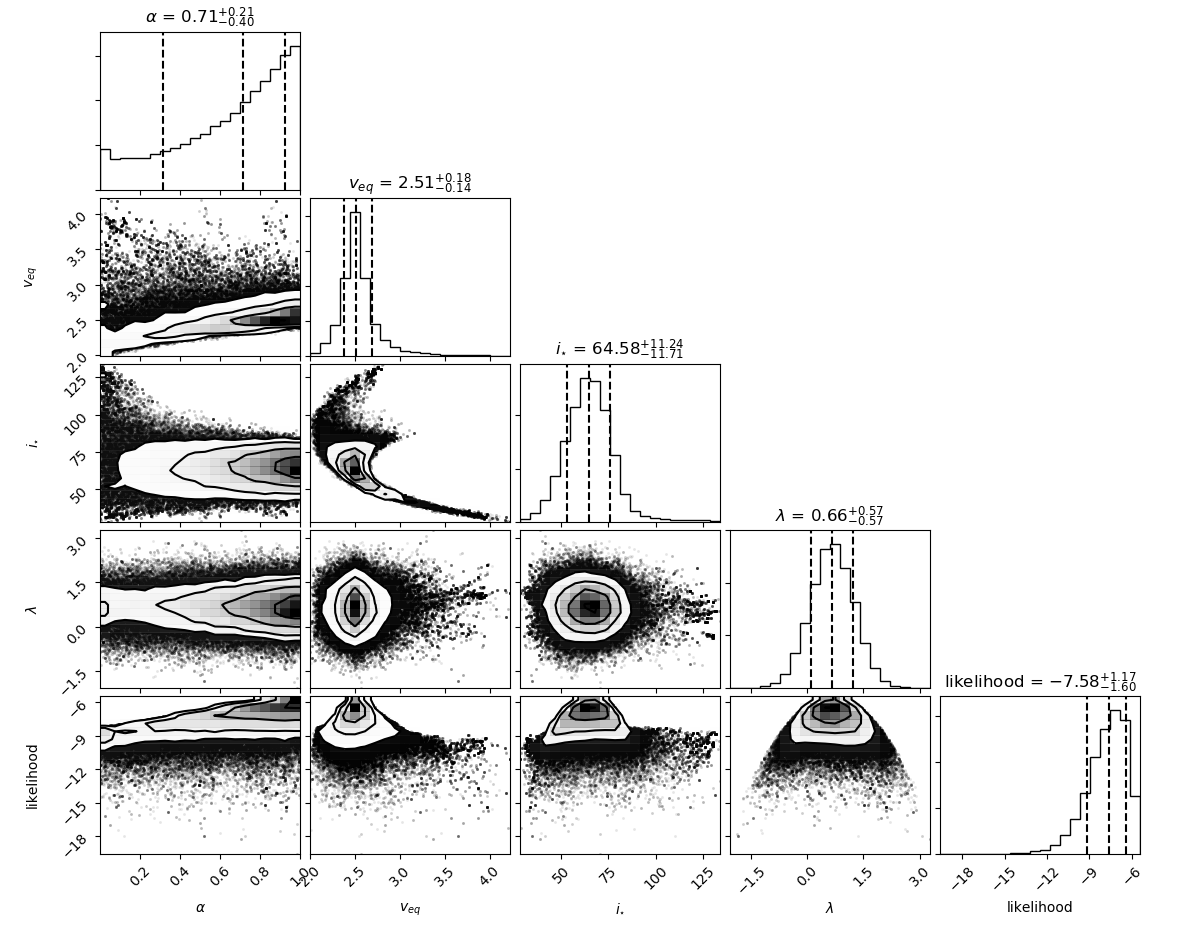}
\caption[]{
WASP-52 corner plot for differential stellar rotation, assuming a positive (solar-like) rotational shear (DR, $0 \le \alpha \le 1$); displayed are the one and two dimensional projections of the posterior probability distributions and their corresponding log-normal likelihood estimates. Vertical dashed lines correspond to the median values and their respective 1$\sigma$ uncertainties calculated from the 16th, 50th, and 84th percentiles of the samples.   
} 
\label{fig:wasp-52_dr_solar_mcmc}
\end{figure*}
\end{center}

\begin{center}
\begin{figure*}[t!]
\centering
\includegraphics[trim=0.cm 0.cm 0.cm 0.cm, clip, scale=0.63]{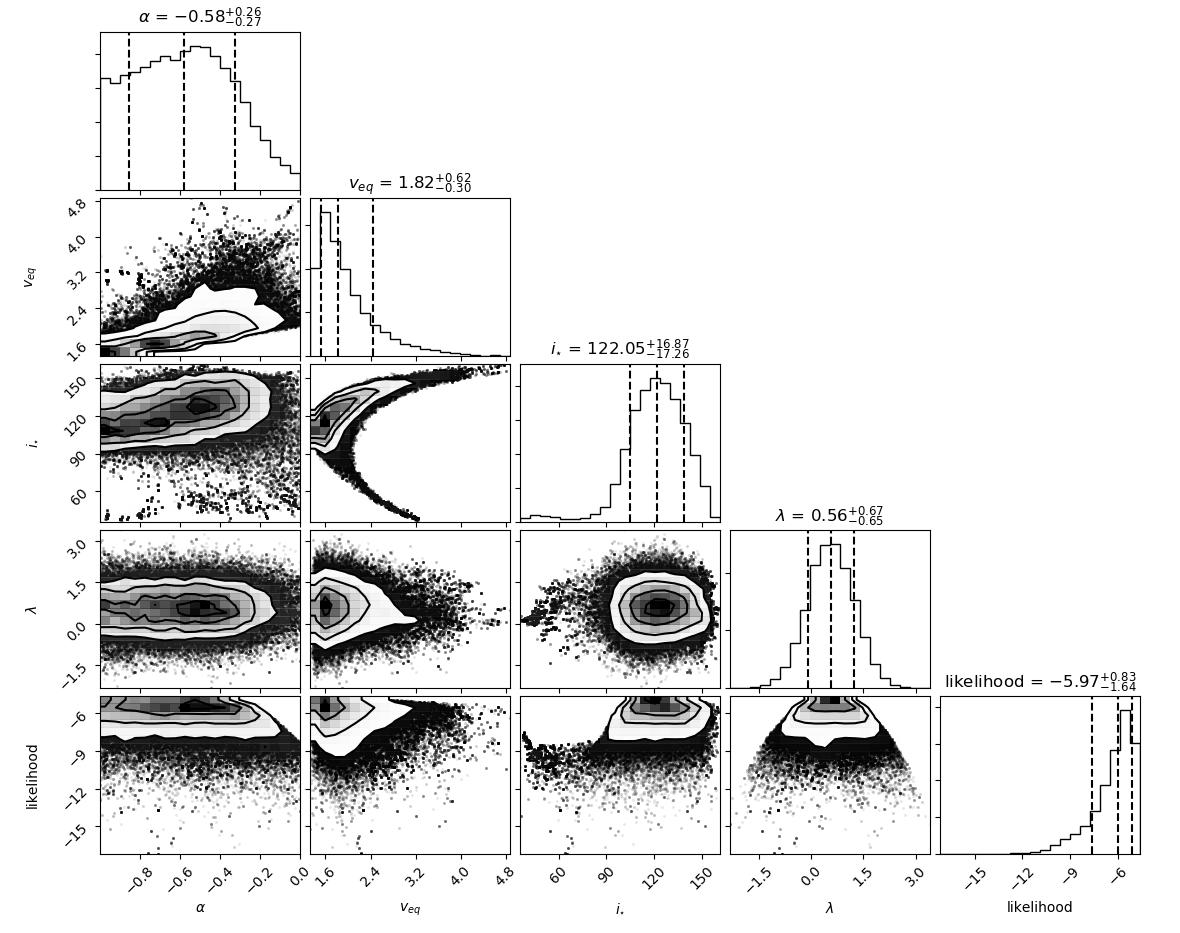}
\caption[]{
WASP-52 corner plot for differential stellar rotation, assuming a negative (antisolar) rotational shear (DR, $-1 \le \alpha  \le 0$); displayed are the one and two dimensional projections of the posterior probability distributions and their corresponding log-normal likelihood estimates. Vertical dashed lines correspond to the median values and their respective 1$\sigma$ uncertainties calculated from the 16th, 50th, and 84th percentiles of the samples.   
} 
\label{fig:wasp-52_dr_antisolar_mcmc}
\end{figure*}
\end{center}

\begin{center}
\begin{figure*}[t!]
\centering
\includegraphics[trim=0.cm 0.cm 0.cm 0.cm, clip, scale=0.63]{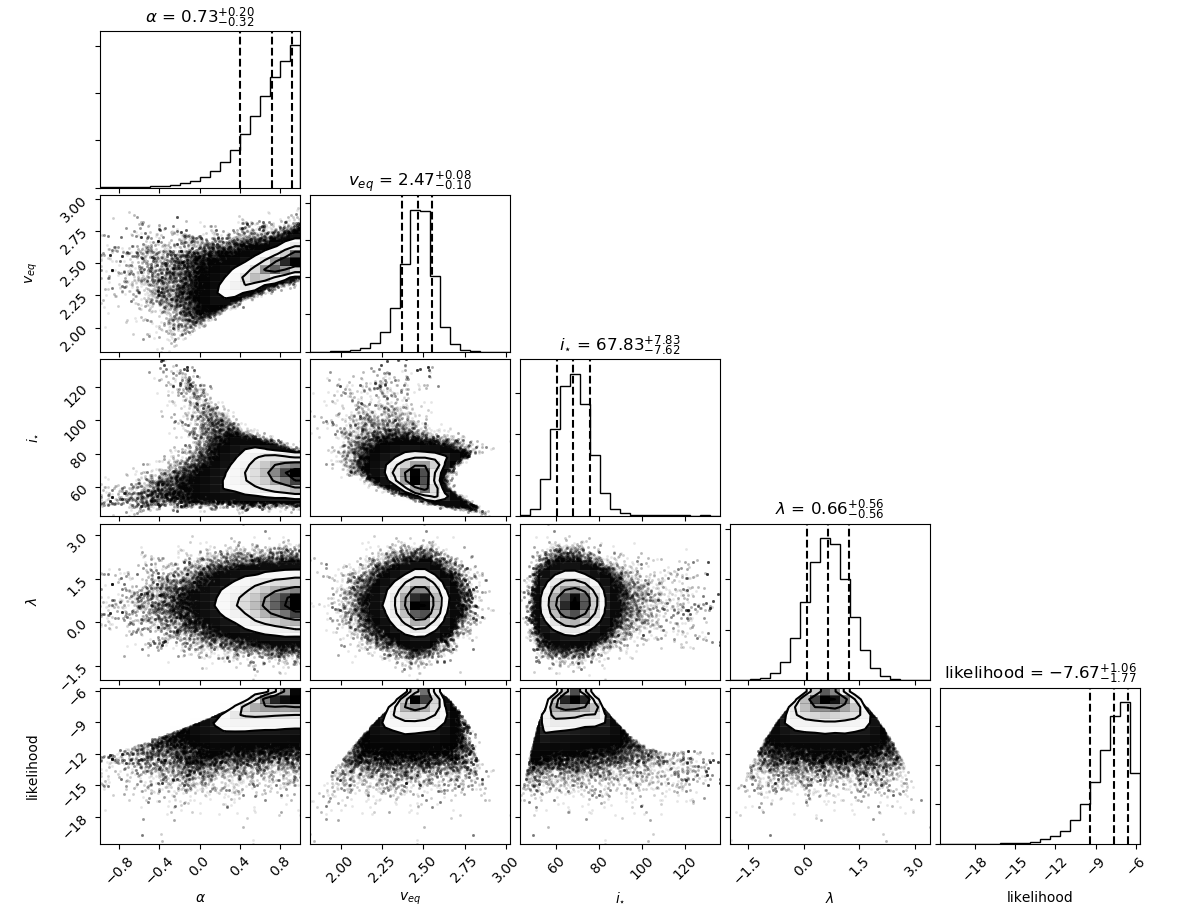}
\caption[]{
WASP-52 corner plot for differential stellar rotation, with priors on $v_{eq}$ and $i_{\star}$ from long-term photometric monitoring (DR, phot. prior); displayed are the one and two dimensional projections of the posterior probability distributions and their corresponding log-normal likelihood estimates. Vertical dashed lines correspond to the median values and their respective 1$\sigma$ uncertainties calculated from the 16th, 50th, and 84th percentiles of the samples.   
} 
\label{fig:wasp-52_dr_soap_mcmc}
\end{figure*}
\end{center}

\section{Posterior Distributions for HAT-P-30}
\label{apn:mcmc_hatp30}

\begin{center}
\begin{figure*}[t!]
\centering
\includegraphics[trim=0.cm 0.cm 0.cm 0.cm, clip, scale=0.63]{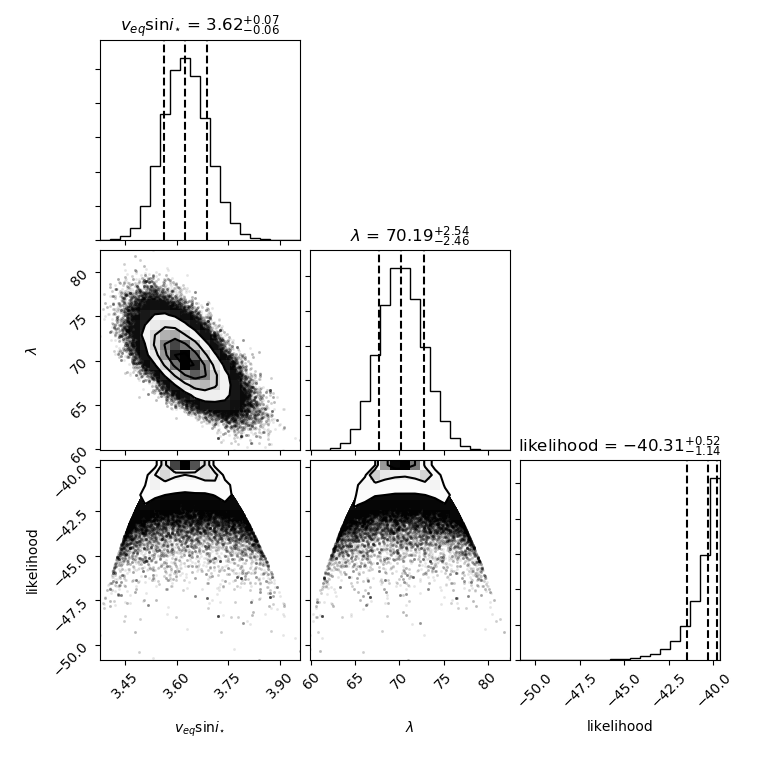}
\caption[]{
HAT-P-30 corner plot for solid body stellar rotation (SB); displayed are the one and two dimensional projections of the posterior probability distributions and their corresponding log-normal likelihood estimates. Vertical dashed lines correspond to the median values and their respective 1$\sigma$ uncertainties calculated from the 16th, 50th, and 84th percentiles of the samples.   
} 
\label{fig:hatp30_sb_mcmc}
\end{figure*}
\end{center}

\begin{center}
\begin{figure*}[t!]
\centering
\includegraphics[trim=0.cm 0.cm 0.cm 0.cm, clip, scale=0.63]{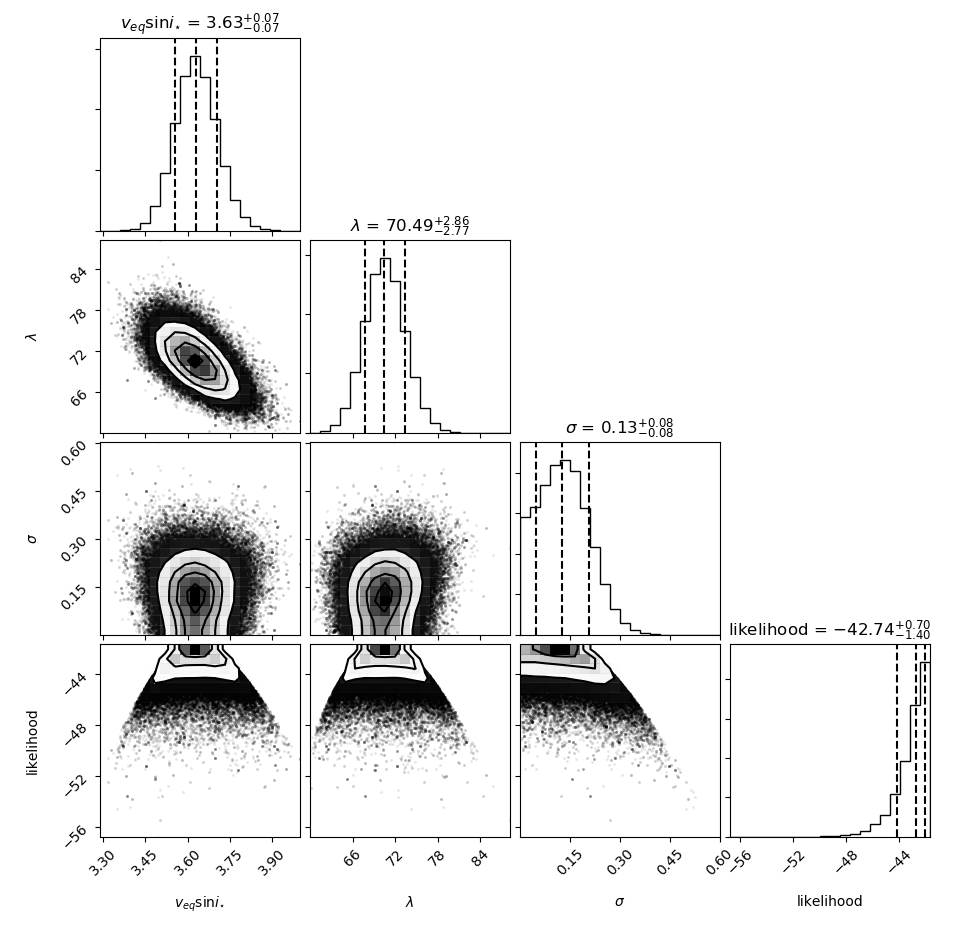}
\caption[]{
HAT-P-30 corner plot for solid body stellar rotation with an additional white noise term (SB+$\sigma$); displayed are the one and two dimensional projections of the posterior probability distributions and their corresponding log-normal likelihood estimates. Vertical dashed lines correspond to the median values and their respective 1$\sigma$ uncertainties calculated from the 16th, 50th, and 84th percentiles of the samples.   
} 
\label{fig:hatp30_sb_jitter_mcmc}
\end{figure*}
\end{center}

\begin{center}
\begin{figure*}[t!]
\centering
\includegraphics[trim=0.cm 0.cm 0.cm 0.cm, clip, scale=0.63]{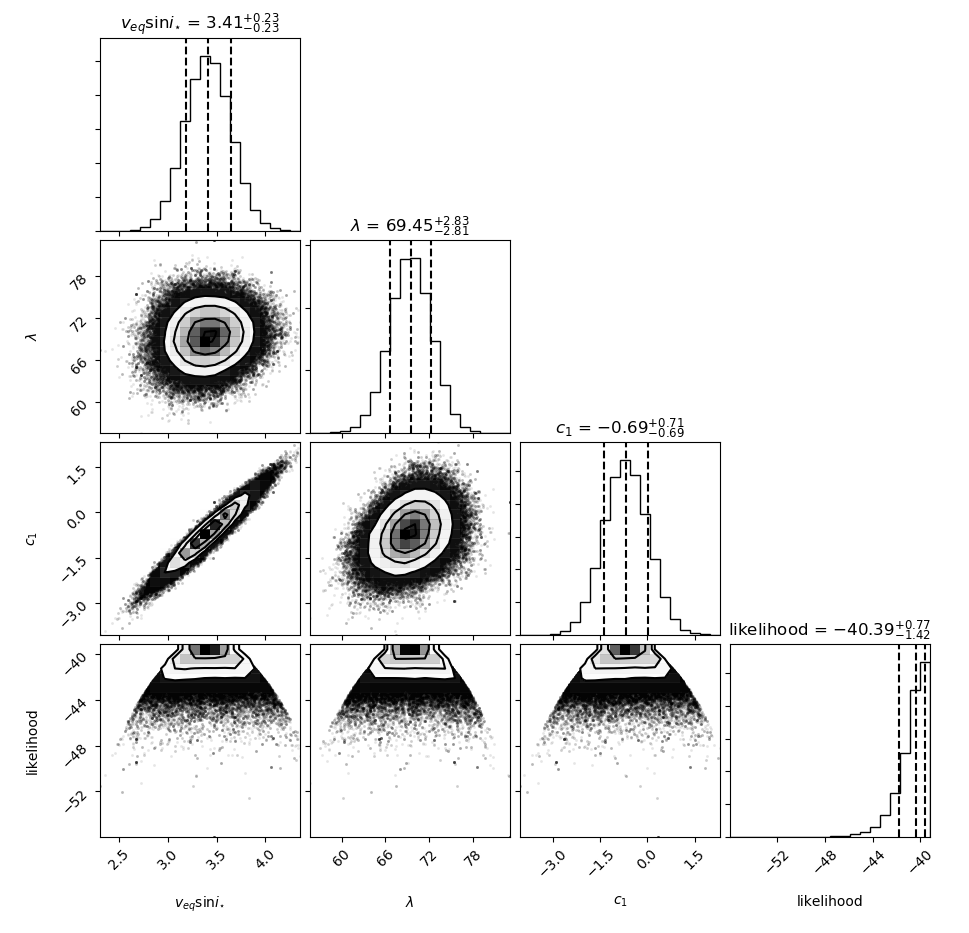}
\caption[]{
HAT-P-30 corner plot for solid body stellar rotation with a linear centre-to-limb variation due to convection (SB+CLV$_{\rm{lin}}$); displayed are the one and two dimensional projections of the posterior probability distributions and their corresponding log-normal likelihood estimates. Vertical dashed lines correspond to the median values and their respective 1$\sigma$ uncertainties calculated from the 16th, 50th, and 84th percentiles of the samples.  
} 
\label{fig:hatp30_sb_cb_lin_mcmc}
\end{figure*}
\end{center}

\begin{center}
\begin{figure*}[t!]
\centering
\includegraphics[trim=0.cm 0.cm 0.cm 0.cm, clip, scale=0.63]{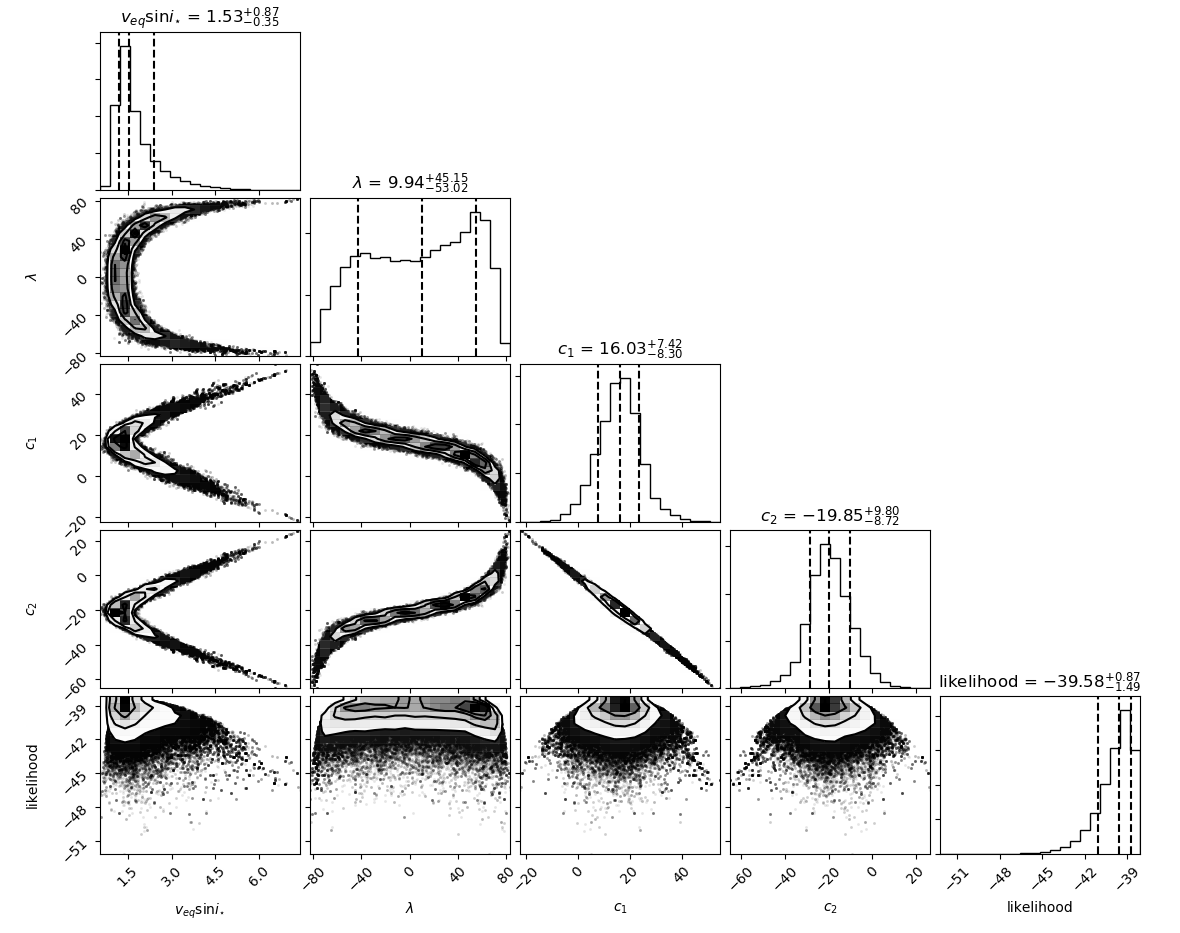}
\caption[]{
HAT-P-30 corner plot for solid body stellar rotation with a quadratic centre-to-limb variation due to convection (SB+CLV$_{\rm{quad}}$); displayed are the one and two dimensional projections of the posterior probability distributions and their corresponding log-normal likelihood estimates. Vertical dashed lines correspond to the median values and their respective 1$\sigma$ uncertainties calculated from the 16th, 50th, and 84th percentiles of the samples. 
} 
\label{fig:hatp30_sb_cb_quad_mcmc}
\end{figure*}
\end{center}

\begin{center}
\begin{figure*}[t!]
\centering
\includegraphics[trim=0.cm 0.cm 0.cm 0.cm, clip, scale=0.63]{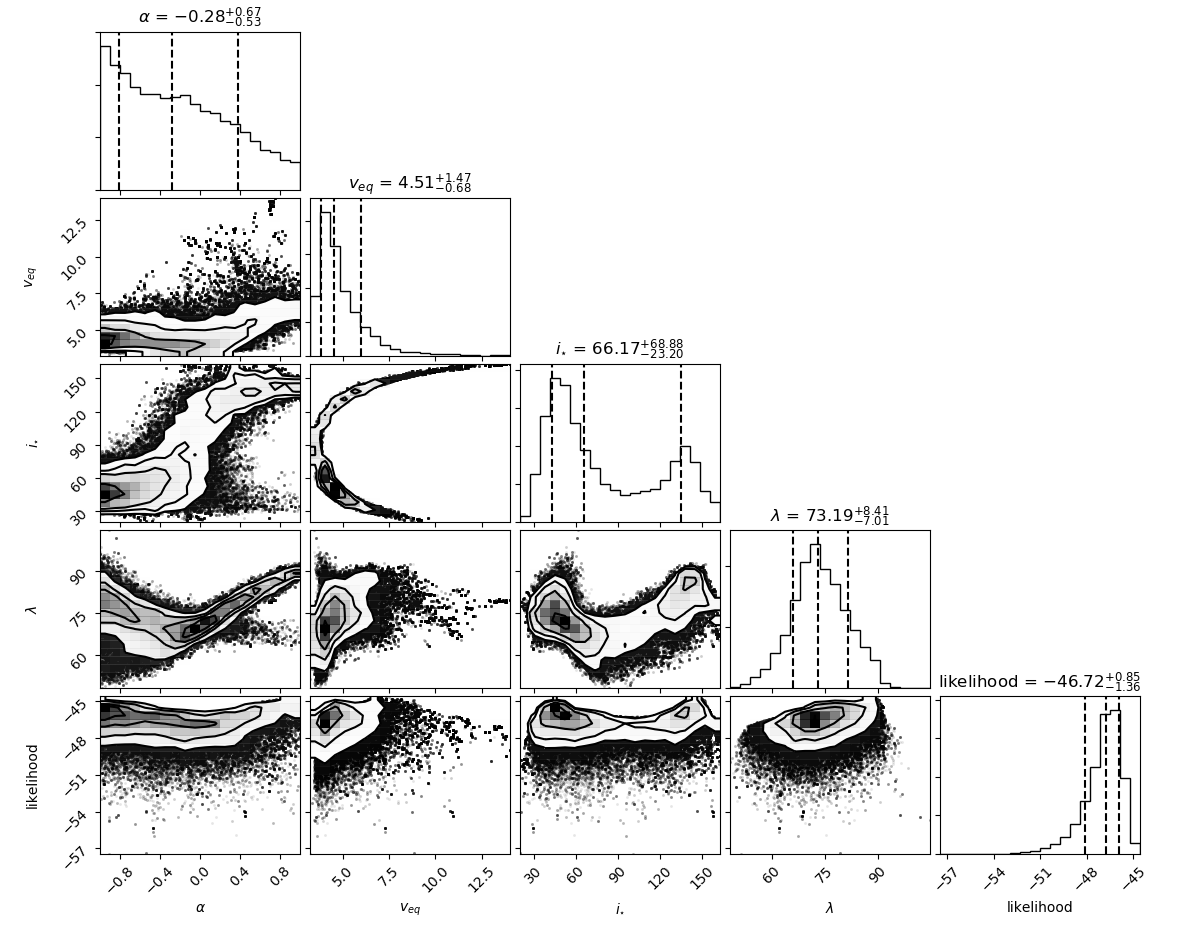}
\caption[]{
HAT-P-30 corner plot for differential stellar rotation (DR); displayed are the one and two dimensional projections of the posterior probability distributions and their corresponding log-normal likelihood estimates. Vertical dashed lines correspond to the median values and their respective 1$\sigma$ uncertainties calculated from the 16th, 50th, and 84th percentiles of the samples.   
} 
\label{fig:hatp30_dr_mcmc}
\end{figure*}
\end{center}

\begin{center}
\begin{figure*}[t!]
\centering
\includegraphics[trim=0.cm 0.cm 0.cm 0.cm, clip, scale=0.63]{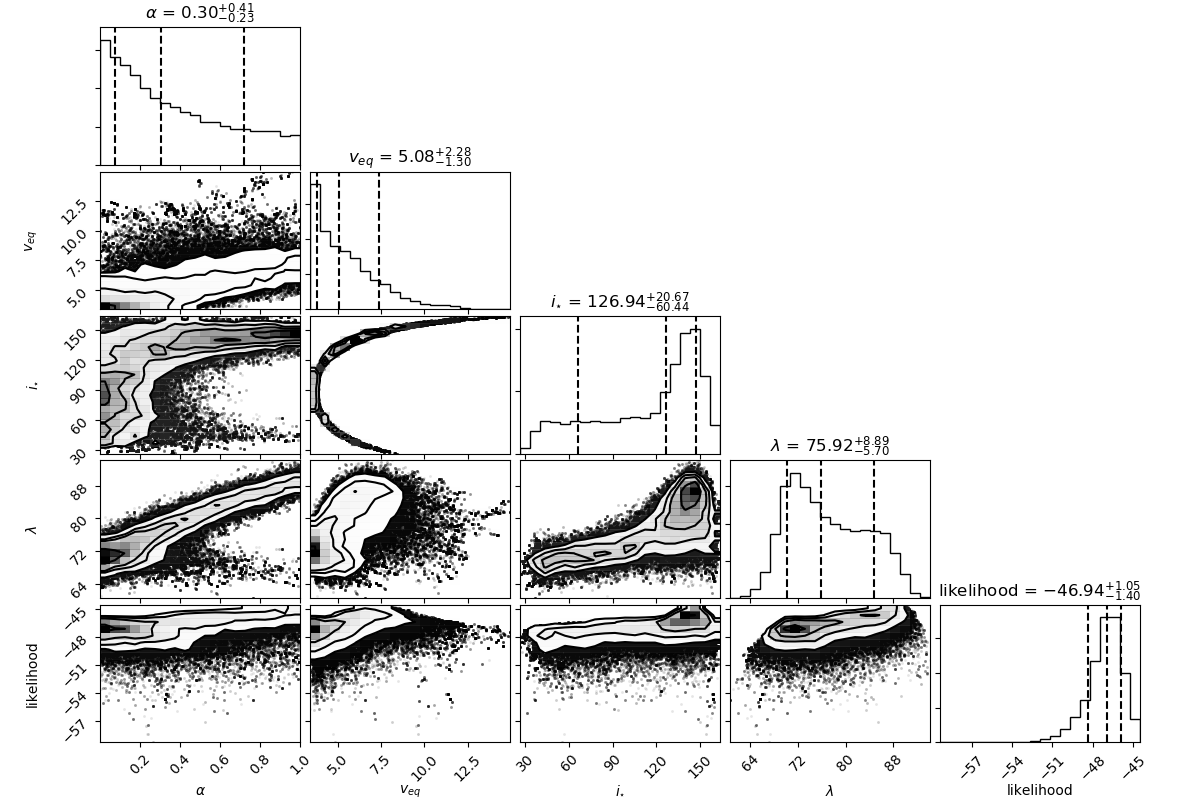}
\caption[]{
HAT-P-30 corner plot for differential stellar rotation, assuming a positive (solar-like) rotational shear (DR, $0 \le \alpha \le 1$); displayed are the one and two dimensional projections of the posterior probability distributions and their corresponding log-normal likelihood estimates. Vertical dashed lines correspond to the median values and their respective 1$\sigma$ uncertainties calculated from the 16th, 50th, and 84th percentiles of the samples.   
} 
\label{fig:hatp30_dr_solar_mcmc}
\end{figure*}
\end{center}

\begin{center}
\begin{figure*}[t!]
\centering
\includegraphics[trim=0.cm 0.cm 0.cm 0.cm, clip, scale=0.63]{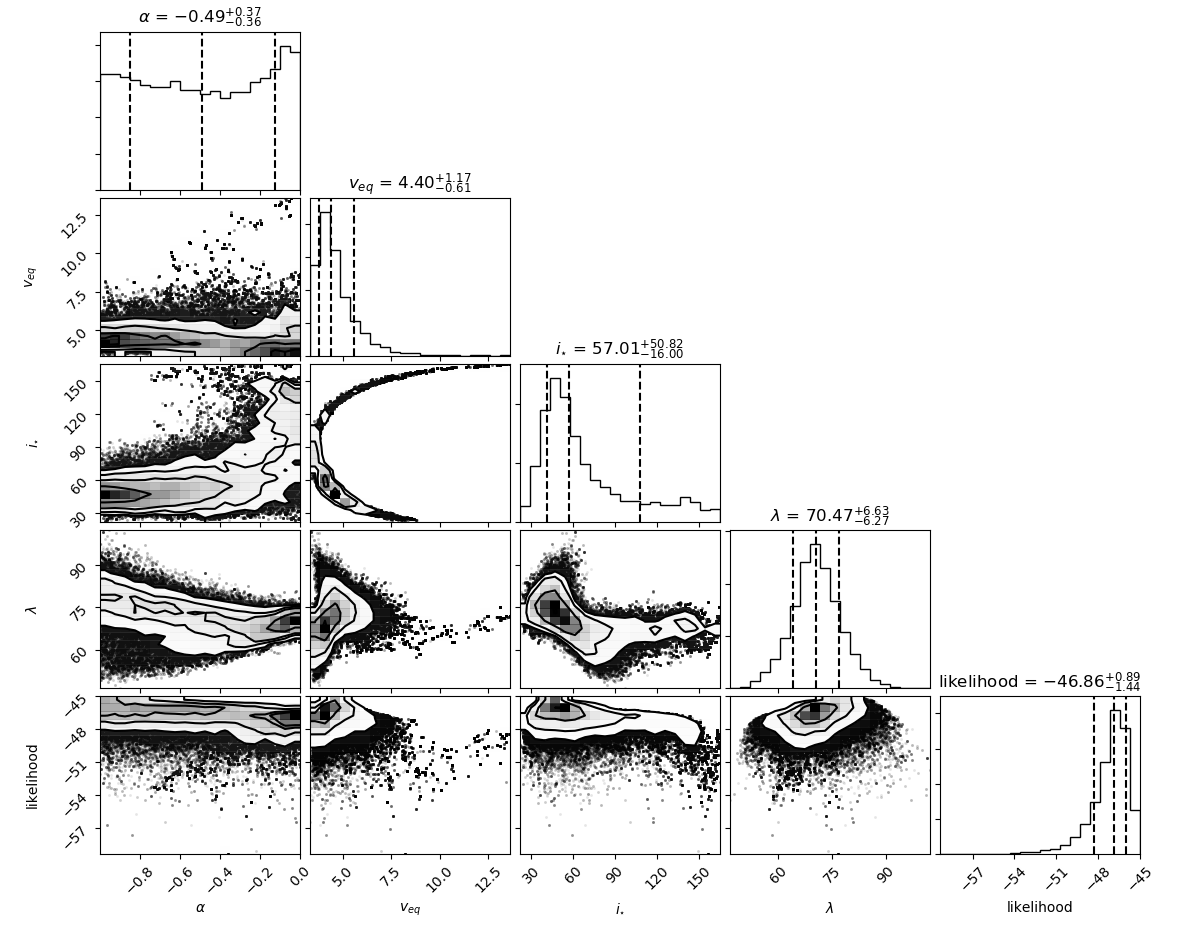}
\caption[]{
HAT-P-30 corner plot for differential stellar rotation, assuming a negative (antisolar) rotational shear (DR, $-1 \le \alpha \le 0$); displayed are the one and two dimensional projections of the posterior probability distributions and their corresponding log-normal likelihood estimates. Vertical dashed lines correspond to the median values and their respective 1$\sigma$ uncertainties calculated from the 16th, 50th, and 84th percentiles of the samples. 
} 
\label{fig:hatp30_dr_antisolar_mcmc}
\end{figure*}
\end{center}

\begin{center}
\begin{figure*}[t!]
\centering
\includegraphics[trim=0.cm 0.cm 0.cm 0.cm, clip, scale=0.63]{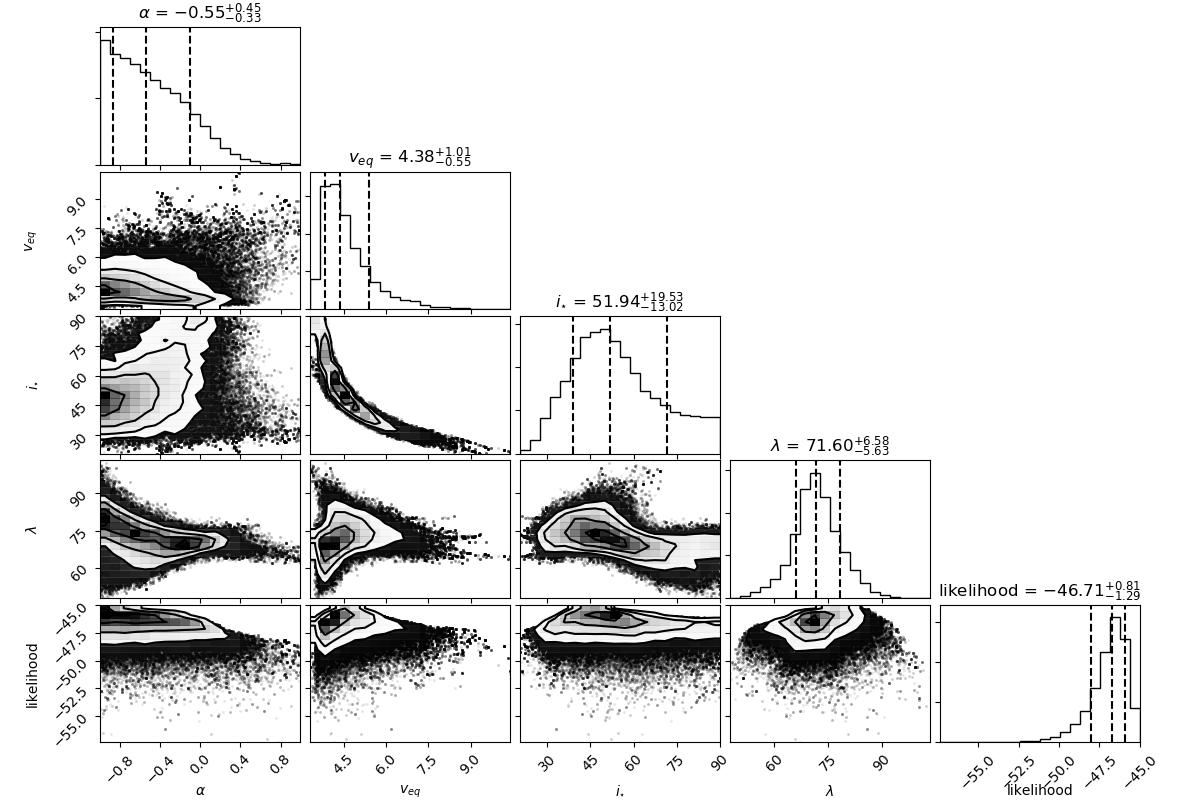}
\caption[]{
HAT-P-30 corner plot for differential stellar rotation, assuming a stellar inclination pointing towards the line-of-sight (DR, $0 \le i_{\star} \le 90$); displayed are the one and two dimensional projections of the posterior probability distributions and their corresponding log-normal likelihood estimates. Vertical dashed lines correspond to the median values and their respective 1$\sigma$ uncertainties calculated from the 16th, 50th, and 84th percentiles of the samples. 
} 
\label{fig:hatp30_dr_lt90_mcmc}
\end{figure*}
\end{center}

\begin{center}
\begin{figure*}[t!]
\centering
\includegraphics[trim=0.cm 0.cm 0.cm 0.cm, clip, scale=0.63]{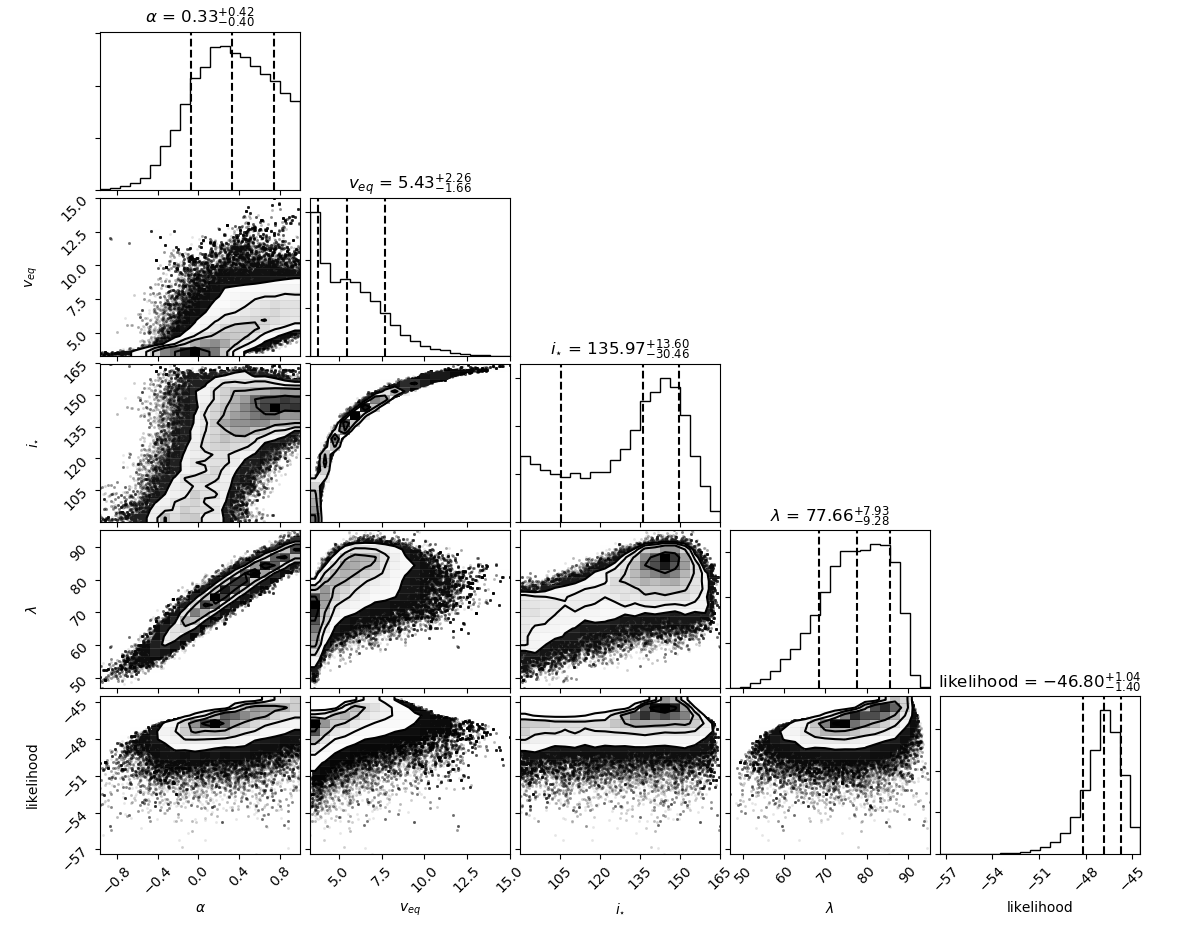}
\caption[]{
HAT-P-30 corner plot for differential stellar rotation, assuming a stellar inclination pointing away from the line-of-sight (DR, $90 \le i_{\star} \le 180$); displayed are the one and two dimensional projections of the posterior probability distributions and their corresponding log-normal likelihood estimates. Vertical dashed lines correspond to the median values and their respective 1$\sigma$ uncertainties calculated from the 16th, 50th, and 84th percentiles of the samples. 
} 
\label{fig:hatp30_dr_gt90_mcmc}
\end{figure*}
\end{center}

\end{appendix} 

\end{document}